\documentclass[12pt,preprint,url]{aastex}
\usepackage{color}
\newcommand{\hl}[1]{{\bf #1}}

\usepackage{amsbsy}
\usepackage{amsmath}
\newcommand{\be}{\begin{equation}}
\newcommand{\ee}{\end{equation}}

\def\ltsima{$\; \buildrel < \over \sim \;$}

\def\lsim{\lower.5ex\hbox{\ltsima}}
\def\gtsima{$\; \buildrel > \over \sim \;$}
\def\gsim{\lower.5ex\hbox{\gtsima}}
\shorttitle{Predicting pulsar glitch epochs}
\shortauthors{Melatos et al.}

\begin{document}
\title{Pulsar glitch activity as a state-dependent Poisson process:
 parameter estimation and epoch prediction}

\author{A. Melatos\altaffilmark{1,2} 
 and L. V. Drummond\altaffilmark{1,2}}

\email{amelatos@unimelb.edu.au}

\altaffiltext{1}{School of Physics, University of Melbourne,
 Parkville, VIC 3010, Australia}

\altaffiltext{2}{Australian Research Council Centre of Excellence
 for Gravitational Wave Discovery (OzGrav)}

\begin{abstract}
\noindent 
Rotational glitches in some rotation-powered pulsars 
display power-law size and exponential waiting time distributions.
These statistics are consistent with a state-dependent Poisson process,
where the glitch rate is an increasing function of a
global stress variable (e.g.\ crust-superfluid angular velocity lag),
diverges at a threshold stress,
increases smoothly while the star spins down, 
and decreases step-wise at each glitch.
A minimal, seven-parameter, maximum likelihood model is calculated
for PSR J1740$-$3015, PSR J0534$+$2200, and PSR J0631$+$1036,
the three objects with the largest samples whose
glitch activity is Poisson-like.
The estimated parameters have theoretically reasonable values
and contain useful information about the glitch microphysics.
It is shown that the maximum likelihood,
state-dependent Poisson model is a marginally (23--27 per cent) better 
{\em post factum} ``predictor'' of historical glitch epochs
than a homogeneous Poisson process for PSR J1740$-$3015 and PSR J0631$+$1036
and a comparable predictor for PSR J0534$+$2200.
Monte Carlo simulations imply that $\gtrsim 50$ glitches
are needed to test reliably whether one model outperforms the other.
It is predicted that the next glitch will occur at 
Modified Julian Date (MJD) $57784\pm 256.8$,
$60713 \pm 1935$, and $57406 \pm 1444$ for the above three objects respectively.
The analysis does not apply to quasiperiodic glitchers like
PSR J0537$-$6910 and PSR J0835$-$4510,
which are not described accurately by the state-dependent
Poisson model in its original form.
\end{abstract}


\section{Introduction 
 \label{sec:pre1}}
The secular, electromagnetic braking of some rotation-powered pulsars
is interrupted by random, impulsive, spin-up events called glitches
\citep{lyn12}.
Two categories of glitch activity have been identified
\citep{mel08,esp11,onu16,how18,car19b,fue19}:
Poisson-like, in which the waiting time probability density function (PDF)
is exponential,
and the size PDF is a power law over $\lesssim 4\,{\rm dex}$;
and quasiperiodic,
where both the size and waiting time PDFs are approximately Gaussian
and centered on characteristic values.
Most glitching pulsars with statistically significant glitch samples
fall into one of the categories above,
although there is some cross-over.
For example, PSR J0534$+$2200 is Poisson-like
\citep{won01},
PSR J0537$-$6910 is quasiperiodic
\citep{mid06,fer18},
and PSR J1341$-$6220 appears to be a hybrid
\citep{how18}
with log-normal characteristics
\citep{fue19}.
The physical origin of glitches remains unknown but is linked commonly
to a combination of internal processes such as starquakes and
superfluid vortex avalanches
\citep{has15}.

Glitches are stochastic,
in the sense that the epochs and sizes of individual events are unpredictable.
Among the $\gtrsim 200$ pulsars in which glitches have been recorded,
only a handful offer exceptions to this rule.
A tight, three-sigma correlation of $6.5\,{\rm days\,}\mu{\rm Hz}^{-1}$ 
is observed between sizes and forward waiting times in PSR J0537$-$6910,
which can be exploited to predict the epoch of the next glitch
to within $\pm 3$ days;
see the `staircase plots' in Fig.\ 8 in \citet{mid06}
and Fig.\ 2 in \citet{fer18}.
An analogous three-sigma correlation is observed in PSR J1801$-$2304
\citep{mel18,fue19}.
Weaker correlations between sizes and forward waiting times
can also be discerned in PSR J1341$-$6220
\citep{yua10,mel18,fue19},
PSR J0205$+$6449
\citep{mel18,fue19},
and
PSR J1645$-$0317 
\citep{sha09}.
However more data are needed to evaluate their power as an epoch prediction tool,
and the events in PSR J1645$-$0317 are `slow' glitches,
whose rise times are resolved over $\sim 1\,{\rm yr}$.
Beyond the above examples, no size-waiting-time correlations have been measured,
which could be exploited for the purpose of epoch prediction,
not even in quasiperiodic objects.
Independently, in PSR J0835$-$4510,
there is evidence that a glitch is triggered,
when the transient impulse response to the previous glitch recovers fully,
if one corrects for the quasiexponential post-glitch relaxation
and the inter-glitch frequency second derivative
\citep{akb17}.
This approach predicts glitch epochs in PSR J0835$-$4510
with an uncertainty of about $\pm 150\,{\rm days}$
and deserves to be tested against other pulsars, e.g.\ \citet{yu13}.
Second-frequency-derivative corrections can be related to 
the physics of nonlinear vortex creep
\citep{alp84}.

In this paper, we approach the challenge of epoch prediction 
by modeling glitch activity as a state-dependent Poisson process
\citep{ful17,mel18,car19b,car19c},
in which the glitch rate is variable and depends instantaneously
on the `stress' in the system
(elastic stress in the starquake picture,
crust-superfluid differential rotation in the vortex avalanche picture).
The model naturally predicts two categories of glitch activity
\citep{ful17,car19b}
and a paucity of size-waiting-time auto- and cross-correlations
\citep{mel18,car19c}
in line with observations.
It also allows us to reconstruct the stress history of a pulsar
from the observed sequence of glitch sizes and epochs
in terms of seven model parameters
and hence derive a maximum likelihood estimate of the stress value today,
which in turn predicts a Poisson waiting time to the next glitch.
This prediction should be better than the unmodeled prediction
based on the time-averaged rate inferred from the waiting time PDF,
if the stress and hence the instantaneous rate vary
from one glitch to the next
\citep{ful17,mel18,car19a}.

The paper is structured as follows.
The minimal version of the state-dependent Poisson model is reviewed briefly
in \S\ref{sec:pre2},
and a maximum likelihood estimator is constructed
for the seven parameters which control the model's behavior.
Maximizing the likelihood is not trivial.
Two numerical maximization techniques,
particle swarm and nested sampling,
are presented and tested for accuracy against synthetic data
in \S\ref{sec:pre3}.
The seven model parameters are estimated for three objects
with Poisson-like glitch activity in \S\ref{sec:pre4},
namely PSR J1740$-$3015, PSR J0534$+$2200, and PSR J0631$+$1036.
The physical significance of the best-fit parameters is discussed.
Finally the maximum likelihood estimates in \S\ref{sec:pre4}
are used to ``predict'' {\em post factum} the epochs of past glitches
for the above three objects (as a validation test)
and then to genuinely predict the epoch of the next, future glitch
(with confidence intervals).
The predictions are therefore directly falsifiable over time.
Parameter estimation for other, Poisson-like glitchers 
with smaller samples is premature at this juncture.
The minimal version of the state-dependent Poisson model 
introduced by \citet{ful17}
does not describe accurately the glitch activity of quasiperiodic
glitchers like PSR J0537$-$6910 and PSR J0835$-$4510,
so the latter category of object is not analysed here.

\section{State-dependent Poisson process
 \label{sec:pre2}}
For various physical mechanisms, glitch activity is controlled
in the mean-field approximation by a single, global, random variable,
$X(t)$, which measures the spatially averaged crust-superfluid 
angular velocity lag (vortex avalanche picture)
or the crustal elastic stress (starquake picture)
as a function of time $t$.
In between glitches, $X(t)$ increases secularly in response
to the electromagnetic braking torque $N_{\rm em}$,
with $\dot{X} \propto |N_{\rm em}|$.
When a glitch occurs, $X$ decreases discontinuously
by a random amount determined by the avalanche physics governing
the glitch trigger,
e.g.\ vortex unpinning \citep{war11}
or crust cracking \citep{chu10b}.
The combination of a slow, global driver,
which adds stress,
and local, stick-slip relaxation,
which releases stress,
is characteristic of systems that exhibit self-organized criticality
\citep{jen98,mel08}.

In \S\ref{sec:pre2a} and \S\ref{sec:pre2b},
we model the evolution of $X(t)$ in an idealized fashion 
as a state-dependent Poisson process
\citep{dal07,whe08,war13,ful17,car19b}.
We then construct a maximum likelihood estimator of the model's
seven control parameters from an observed sequence of glitch sizes
and waiting times in \S\ref{sec:pre2c}.
A state-dependent Poisson process occupies a well-defined place
within the established taxonomy of stochastic processes.
In a general sense, it is a marked renewal process:
marked, because every event is tagged with auxiliary information
(here, the glitch size) besides its epoch;
and renewal, because it involves stochastically recurring events,
whose waiting time distribution resets after every glitch.
A state-dependent Poisson process may also be regarded as doubly stochastic,
because the event rate itself is a random variable;
it increases deterministically between glitches,
but its starting value immediately after every glitch depends on 
the random, post-glitch value of $X$.
A state-dependent Poisson process is more general than 
an inhomogeneous, Poisson process because it is marked,
and because the rate is stochastic.
It is also more general than a continuous-time jump Markov process,
because the waiting time distribution depends on the whole
time series $X(t)$ since the previous glitch,
not just $X$ at the previous time step.
(The glitch sizes constitute a jump Markov process when viewed in isolation.)
The reader is referred to the textbooks by \citet{kin93} and \citet{del17}
for a comprehensive classification of stochastic processes,
including formal definitions of the terms above.

\subsection{Equations of motion
 \label{sec:pre2a}}
It is convenient to write the equations of motion
in dimensionless form in order to identify clearly the
minimum set of irreducible parameters in the model.
Let $X_{\rm c}$ be the critical crust-superfluid angular velocity lag, 
at which the Magnus force exceeds the vortex pinning force globally,
and a glitch is certain to occur
\citep{lin91}.
(An analogous threshold is easy to define in the starquake picture,
but we focus here on glitches triggered by vortex unpinning
for the sake of definiteness.)
Let $I_{\rm c}$ be the moment of inertia of the crust,
i.e.\ the nonsuperfluid stellar component which experiences $N_{\rm em}$ directly,
and let $I_{\rm s}$ be the moment of inertia of the rest of the star.
If we measure the global stress in units of $\tilde{X}_0 = X_{\rm c}$
and time in units of $\tilde{t}_0 = X_{\rm c} I_{\rm c} / N_{\rm em}$,
then the crust-superfluid angular velocity lag satisfies a stochastic
equation of motion given in dimensionless form by
\citep{ful17}
\begin{equation}
  X(t)
 =
 X(T_1^+) + t - T_1 - \gamma \sum_{i=1}^{N(t)} \Delta\nu_i~,
\label{eq:pre1}
\end{equation}
with $\gamma = 2\pi ( 1 + I_{\rm s}/I_{\rm c} )$.
In (\ref{eq:pre1}), $\Delta\nu_i$ is the observed size of the $i$-th glitch,
i.e.\ the $i$-th positive spin frequency jump measured in units of $\tilde{X}_0$,
$T_i$ is the epoch of the $i$-th glitch measured in units of $\tilde{t}_0$,
and $N(t)$ is the number of glitches from $t=0$ 
up to but not including the instant $t$.
We introduce the notation $T_i^+$ ($T_i^-$) to denote the instant
infinitesimally after (before) $T_i$,
so that one has $N(T_{i+1}^-) = N(T_i^+)$ and 
$N(T_{i+1}^+) = N(T_{i+1}^-)+1$.
Equation (\ref{eq:pre1}) contains two random variables, 
$N(t)$ and $\Delta\nu_i$.

Let us assume that glitch triggering is a state-dependent, Poisson process,
whose dimensionless rate function $\lambda(X)$
(i.e.\ the mean number of trigger events per unit time)
increases monotonically with $X$ and diverges as $X\rightarrow X_{\rm c}=1$.
As $X(t)$ evolves deterministically between glitches,
with $\dot{X}(t)=1$, we can apply the standard theory of a variable-rate
Poisson process to write down the waiting time PDF,
\begin{equation}
 p[T_i - T_{i-1} | X(T_{i-1}^+) ]
 =
 \lambda[ X(T_i^-) ]
 \exp\left\{
  - \int_{T_{i-1}}^{T_i} dt' \,
  \lambda[ X(t') ]
 \right\}~.
\label{eq:pre2}
\end{equation}
The PDF $p[T_i - T_{i-1} | X(T_{i-1}^+) ]$ is
conditional on the stress $X(T_{i-1}^+)$ just after the previous glitch,
which determines $X(t')$ for $T_{i-1}^+ \leq t' \leq T_i^-$
deterministically and hence $T_i - T_{i-1}$ statistically.
The output of the model does not depend sensitively on the functional form
of $\lambda(X)$; see footnote 6 in \citet{ful17}.
In this paper, for the sake of definiteness, we take
\begin{equation}
 \lambda(X) = \frac{\alpha}{1-X}
\label{eq:pre3}
\end{equation}
with
\begin{equation}
 \alpha = \frac{X_{\rm c} I_{\rm c} \lambda_0}{N_{\rm em}}~,
\end{equation}
where $\lambda_0$ is a reference trigger rate
(with dimensions of inverse time) 
equal to half the mean number of glitch triggers per unit time for $X=1/2$.
By fitting the model to data, as in \S\ref{sec:pre4} and \S\ref{sec:pre5},
we can extract $\alpha$ and $\tilde{t}_0$ and hence infer
$\lambda_0 = \alpha \tilde{t}_0^{-1}$,
garnering an important clue about the trigger physics,
e.g.\ vortex unpinning or crust cracking
\citep{war11,chu10b}.

In the state-dependent Poisson process modeled by \citet{ful17},
glitch sizes are uncorrelated with the accumulated stress,
except that we have $\gamma\Delta\nu_i \leq X(T_i^-)$ for all $i$
to keep $X(t)$ nonnegative.
For example, the probability of a relatively large glitch
does not increase, as $X(t)$ approaches $X_{\rm c}$.
The lack of a $\Delta\nu_i$-$X(T_i^-)$ correlation,
although counterintuitive, 
is a general feature of self-organized critical systems
\citep{jen98}
and is verified by Gross-Pitaevskii simulations of superfluid
vortex avalanches \citep{war11}.
In contrast, glitch sizes do affect the waiting time statistics
through $X(T_{i-1}^+)$ in (\ref{eq:pre2}),
and waiting times affect glitch sizes through the constraint
$\gamma\Delta\nu_i \leq X(T_i^-)$  and (\ref{eq:pre1}),
creating potentially observable correlations between
$\Delta\nu_i$ and $T_{i+1}-T_i$ ($T_i - T_{i-1}$)
in the regime $\alpha \ll 1$ ($\alpha \gg 1$)
\citep{ful17,mel18,car19c}.
Therefore it is advantageous to exploit both the observed sequences
$\Delta\nu_1,\dots,\Delta\nu_N$ and 
$T_1,\dots,T_N$
to estimate the parameters of the model and hence predict glitch epochs.

In this paper, we follow \citet{ful17} and model the PDF of the
avalanche sizes, $X(T_i^-)-X(T_i^+)$, as a power law.
Other functional forms are defensible, of course,
but a power law is consistent with the scale invariant size PDFs
observed in objects with Poisson-like glitches
\citep{mel08,esp14,ash17,how18,sha18,fue19},
which have exponents $\delta$ in the range
$0.4\lesssim \delta \lesssim 2.4$;
see Table 3 and Figure 9 in \citet{mel08}.
It is also consistent with various self-organized critical systems
including earthquakes 
\citep{jen98}
and emerges from Gross-Pitaevskii simulations of superfluid vortex avalanches
\citep{war11,mel15},
although computational cost restricts the dynamic range in the simulations
to $\approx 1.5\,{\rm dex}$.
Letting $\eta[ X(T_i^+) | X(T_i^-)]$ be the PDF of the stress $X(T_i^+)$
immediately after the glitch, conditional on the stress equalling $X(T_i^-)$
immediately before the glitch, we write
\begin{equation}
 \eta[X(T_i^-)-\gamma\Delta\nu_i | X(T_i^-) ]
 =
 \frac{(1-\delta) (\gamma\Delta\nu_i)^{-\delta}}
  {(1-\beta^{1-\delta}) [ X(T_i^-) ]^{1-\delta}}~,
\label{eq:pre5}
\end{equation}
where $\beta$ is the minimum fractional jump size
required for the PDF to be normalizable,
i.e.\ $\gamma\Delta\nu_i \geq \beta X(T_i^-)$.
\footnote{
The conditional jump distribution $\eta(Y|Z)$ is a probability density
in the variable $Y$, normalized as $1=\int dY \, \eta(Y|Z)$.
Equations (18) and (19) in \citet{ful17} involve a Heaviside factor
$\eta(Y|Z) \propto H(Z-Y-\beta Z)$, which keeps the stress nonnegative,
but this factor is always unity in (\ref{eq:pre5}) for any observed glitch,
because an observed glitch necessarily entails a valid $(Y,Z)$ combination.
}
Equation (\ref{eq:pre5}) is derived by assuming
$\eta[X(T_i^+)|X(T_i^-)] = K (\Delta \nu_i)^{-\delta}$
on the phenomenological basis discussed above
and evaluating $K$ by normalizing $\eta$ on the interval
$\beta X(T_i^-) \leq X(T_i^-) - X(T_i^+) \leq X(T_i^-)$, viz.
\begin{equation}
 1 = 
 K \int_{\gamma^{-1}\beta X(T_i^-)}^{\gamma^{-1} X(T_i^-)}
 d(\Delta \nu_i) \, (\Delta \nu_i)^{-\delta}~. 
\end{equation}

\subsection{Critical spin-down rate
 \label{sec:pre2b}}
The statistical behavior of the system described by 
(\ref{eq:pre1})--(\ref{eq:pre5})
divides cleanly into two regimes, 
$\alpha \lesssim \alpha_{\rm c}(\beta)$
and
$\alpha \gtrsim \alpha_{\rm c}(\beta)$,
with 
$\alpha_{\rm c} (\beta) \sim 1$.
The regimes are studied thoroughly in \S{4.4} and \S{5.5} in \citet{ful17};
the reader is referred to the latter reference for a complete discussion.
The PDFs generated by (\ref{eq:pre1})--(\ref{eq:pre5})
can be calculated analytically, 
when $\eta[X(T_i^+) | X(T_i^-)]$ is separable,
and numerically when it is not, as in (\ref{eq:pre5}).

In summary,
in the slow-spin-down regime 
$\alpha \gtrsim \alpha_{\rm c}(\beta)$,
equations (\ref{eq:pre1})--(\ref{eq:pre5}) generate:
(i) a power-law size PDF, whose inertial range increases
as $\beta$ decreases,
and whose mean decreases, as $\alpha$ increases
[Figures 6 and 7 in \citet{ful17}];
(ii) an exponential waiting time PDF,
whose mean increases, as $\alpha$ increases
[Figure 8 in \citet{ful17}];
and (iii) a weak correlation between size and backward waiting time,
which strengthens (but remains weak), as $\alpha$ increases
\citep{mel18}.
These properties are broadly consistent with observations of pulsars
with Poisson-like glitch activity, 
e.g.\ PSR J1740$-$3015, PSR J0534$+$2200, and PSR J0631$+$1036.

In the fast-spin-down regime
$\alpha \lesssim \alpha_{\rm c}(\beta)$,
equations (\ref{eq:pre1})--(\ref{eq:pre5}) generate:
(i) size and waiting time PDFs of the same functional form,
whose inertial ranges depend on $\beta$;
(ii) mean sizes and waiting times which are independent of $\alpha$
but decrease, as $\beta$ decreases;
and (iii) a strong correlation between size and forward waiting time,
which strengthens, as $\alpha$ decreases
\citep{mel18}.
Aspects of these properties are broadly consistent
with observations of quasiperiodic glitchers
like PSR J0537$-$6910 and PSR J0835$-$4510,
e.g.\ the size and waiting time PDFs are similar,
and PSR J0537$-$6910 exhibits a strong size-waiting-time correlation.
Other aspects are inconsistent,
e.g.\ $\eta[X(T_i^+) | X(T_i^-)]$ of the form (\ref{eq:pre5})
leads to power-law size and waiting time PDFs,
whereas observations reveal the PDFs to be approximately Gaussian.
Therefore, as the state-dependent Poisson model in its currently
published form describes some but not all of the properties of
quasiperiodic glitchers accurately,
we do not analyse such objects in this paper.

\subsection{Maximum likelihood estimator
 \label{sec:pre2c}}
The equations of motion in \S\ref{sec:pre2a} can be combined 
to construct a likelihood function,
${\cal L}[\tilde{X}_0,\tilde{t}_0,\alpha,X(T_1^+),\gamma,\beta,\delta
 | T_1,\dots,T_N,\Delta\nu_1,\dots,\Delta\nu_N]$,
which measures the likelihood that the hypothesis consisting of
the state-dependent Poisson process in \S\ref{sec:pre2a}
with parameters
$\theta=\{ \tilde{X}_0,\tilde{t}_0,\alpha,X(T_1^+),\gamma,\beta,\delta \}$
gives rise to the observed data.
We emphasize again that it is preferable to build ${\cal L}$
out of $T_1,\dots,T_N$ and $\Delta\nu_1,\dots,\Delta\nu_N$ ---
not just $T_1,\dots,T_N$ ---
even if one cares only to predict glitch epochs.
In essence, epoch prediction boils down to estimating the
stress history $X(t)$. All the data are informative.
The waiting times certainly communicate information about $X(t)$;
they tend to shorten, as $X(t)$ approaches $X_{\rm c}$.
But so do the sizes; they cannot be too large,
otherwise $X(t)$ turns negative.

The probability of observing a given glitch sequence is given by
the prior probability that the stress equals $X(T_1^+)$
immediately after the first glitch,
viz.\ $\Pr [ X(T_1^+) ]$,
multiplied by the probability that no glitch occurs
from $T_1$ until $T_2$,
multiplied by the probability that an event of size $\Delta\nu_2$ occurs,
and so on, all the way up to the latest, $N$-th glitch:
\begin{equation}
 \Pr( \{ T_i , \Delta\nu_i \}_{i=2}^N )
 =
 \Pr [ X(T_1^+) ]
 \prod_{i=2}^N
 p[T_i - T_{i-1} | X(T_{i-1}^+) ] 
 \eta[ X(T_i^-) - \gamma\Delta\nu_i | X(T_i^-) ]~.
\label{eq:pre7a}
\end{equation}
Interpreting the $N$-fold product as a likelihood,
and taking the natural logarithm for numerical convenience,
we arrive at the expression
\begin{eqnarray}
 \ln {\cal L}
 & = &
 \ln\Pr [ X(T_1^+) ]
 \nonumber \\
 & & 
 + \sum_{i=2}^N
 \left\{
 \ln \alpha + (\alpha - 1) \ln [ 1-X(T_i^-) ]
 - \alpha \ln [ 1 - X(T_{i-1}^+) ] 
 \right\}
 \nonumber \\ 
 & & 
 + \sum_{i=2}^N
 \left\{
 \ln (1-\delta) - \delta \ln (\gamma \Delta\nu_i )
 - \ln(1-\beta^{1-\delta})
 - (1-\delta) \ln X(T_i^-)
 \right\}~.
\label{eq:pre7}
\end{eqnarray}
Equation (\ref{eq:pre7}) is obtained from (\ref{eq:pre1})--(\ref{eq:pre7a})
by integrating (\ref{eq:pre2}) analytically given the deterministic
evolution $X(t) = X(T_{i-1}^+) + t - T_{i-1}^+$
in the inter-glitch interval $T_{i-1}^+ \leq t \leq T_i^-$:
\begin{eqnarray}
 p[T_i - T_{i-1} | X(T_{i-1}^+) ] 
 & = &
 \frac{\alpha}{1-X(T_i^-)}
 \exp\left[
  -\alpha \int_{T_{i-1}}^{T_i} 
  \frac{dt'}{1 - X(T_{i-1}^+) - t' + T_{i-1}}
 \right]
\label{eq:pre7b}
 \\
 & = &
 \frac{\alpha}{1-X(T_i^-)}
 \left[
  \frac{1-X(T_{i-1}^+) - T_{i} + T_{i-1}}{1-X(T_{i-1}^+)}
 \right]^\alpha~. 
\label{eq:pre7c}
\end{eqnarray}
In the remainder of the paper, we seek to maximize ${\cal L}$
with respect to the parameter set $\theta$.

Two assumptions are made when applying (\ref{eq:pre7}) to a real pulsar. 
First, it is taken for granted that the observed glitch sequence
is a complete sample down to some minimum, resolvable glitch size.
Completeness has been tested rigorously in some objects,
e.g.\ by Monte Carlo simulations in PSR J0358$+$5413
\citep{jan06}
and by comparing with the dispersion of the phase residuals 
in PSR J0534$+$2200
[see \S{3.2} in \citet{esp14}].
However there are many objects,
where rigorous testing of completeness remains to be done.
Variable gaps in timing data,
and the continuing, unavoidable role of subjective human interpretation
in the {\sc tempo2}-based glitch finding process
\citep{edw06,hob06,esp11},
argue for caution.
Second, we assume that $N_{\rm em}/I_{\rm c}$ is constant for
$T_1 \leq t \leq T_N$.
This is valid to within $\leq 5\%$ even in the youngest objects
like PSR J0534$+$2200.

The prior, $\Pr X(T_1^+)$, is sampled immediately after the first glitch.
A Poisson process is memoryless, so we could just as well sample the prior
whenever the pulsar was first monitored, at $T_0 < T_1$,
and include $\Delta\nu_1$ in (\ref{eq:pre7})
by extending the sum to $\sum_{i=1}^N$.
We elect not to do so in this paper, because $T_0$ is tricky to pin down
in the literature for some objects with long monitoring histories.
We show in \S\ref{sec:pre4} {\em a posteriori}
that the parameter estimates and epoch predictions do not depend sensitively
on $\Pr X(T_1^+)$.
For the same reason, we are content to adopt a uniform prior
on $\Pr X(T_1^+)$ on the domain $0\leq X(T_1^+) \leq 1$.
Strictly speaking, for the state-dependent Poisson process in \S\ref{sec:pre2a},
$\Pr X(T_1^+)$ peaks at $0.5 \lesssim X(T_1^+) \lesssim 1$;
see Figure 21 in \citet{ful17}.
One can imagine an iterative procedure,
wherein the model's parameters are estimated from the data assuming
a uniform prior,
$\Pr X(T_1^+)$ is updated from \citet{ful17} using the estimated parameters
as inputs, then the parameters are reestimated with the updated prior.
Although feasible in principle, the procedure is difficult,
because $\Pr X(T_1^+)$ cannot be calculated analytically, when
$\eta[ X(T_i^-) - \gamma\Delta\nu_i | X(T_i^-) ]$
takes the power-law form in (\ref{eq:pre5}).

\section{Numerical method
 \label{sec:pre3}}
The likelihood ${\cal L}$ is a function of seven parameters:
$\tilde{X}_0$, $\tilde{t}_0$, $\alpha$ (or equivalently $\lambda_0$),
$X(T_1^+)$ [or equivalently $\tilde{X}(T_1^+)=\tilde{X}_0 X(T_1^+)$], 
$\gamma$, $\beta$, and $\delta$.
Maximizing ${\cal L}$ with respect to these parameters
is a challenging numerical exercise.
Unsupervised approaches do not work;
human intervention is needed to moderate the process on a case-by-case basis,
e.g.\ bounds on some parameters are determined by trial and error.
In this paper, we execute a supervised, iterative, step-by-step recipe,
which combines analytic bounds on some parameters
(see \S\ref{sec:pre3a})
with two independent, automatic, maximization algorithms
(see \S\ref{sec:pre3b})
and human intervention through various safety checks
on the intermediate results.
The recipe and safety checks are described in Appendix \ref{sec:preappa}.

\subsection{Bounding the parameter space
 \label{sec:pre3a}}
The internal logic of the state-dependent Poisson process
imposes constraints on the seven model parameters $\theta$
\citep{ful17}.
The constraints fall into two classes:
those that affect the waiting time statistics via (\ref{eq:pre2}),
namely $\tilde{t}_0$, $\alpha$, and $X(T_1^+)$;
and those that affect the size statistics via (\ref{eq:pre5}),
namely $\tilde{X}_0$, $\gamma$, $\beta$, and $\delta$.

Consider the waiting times first.
In order to make $0 \leq X(t) \leq 1$ hold at all times between glitches,
while $X(t)$ increases deterministically due to spin down,
we must have $T_{i+1} - T_i \leq 1$ for all $i$ and hence
\begin{equation}
 \tilde{t}_0 
 \geq \max_i 
 ( \tilde{t}_0 T_{i+1} - \tilde{t}_0 T_{i} )~, 
\label{eq:pre8}
\end{equation}
where $\tilde{t}_0 T_i$ is the observed epoch
of the $i$-th glitch,
measured in units of Modified Julian Date (MJD).
In other words,
$\tilde{t}_0$ must be greater than the longest waiting time observed.
In addition the state-dependent Poisson process generates
exponentially distributed waiting times in the slow-spin-down regime,
$\alpha \gtrsim \alpha_{\rm c}(\beta) \sim 1$;
see \S{4.4} in \citet{ful17}.
Hence the option exists to inherit the additional constraint
$\alpha \gtrsim 1$,
if one wishes to apply the model to Poisson-like glitchers only.
We do not impose the latter constraint in this paper in order to keep
the likelihood maximization as general as possible
but we find {\em a posteriori} that it is satisfied
for the three objects analysed in \S\ref{sec:pre4}.

Consider the sizes next.
In order to make $0 \leq X(T_i^+) \leq 1$ hold immediately after every glitch,
we must have 
$\beta X(T_i^-) \leq \gamma\Delta\nu_i \leq X(T_i^-) \leq 1$
for all $i$ and hence
\begin{equation}
 \gamma^{-1} \tilde{X}_0 \geq 
 \max_i \tilde{X}_0 \Delta\nu_i~,
\label{eq:pre10}
\end{equation}
where $\tilde{X}_0 \Delta\nu_i$ is the observed size
of the $i$-th glitch, measured in units of Hz.
In other words, $\tilde{X}_0$ must be greater than the largest spin-up event
observed, after adjusting for the crust-superfluid moments-of-inertia ratio
through $\gamma$.
For a typical neutron star,
modern nuclear physics calculations including entrainment predict
\citep{lin99,and12,cha13,pie14,eya17}
\begin{equation}
 10^{-2} \lesssim 
 -1 + \frac{\gamma}{2\pi} = \frac{I_{\rm s}}{I_{\rm c}} 
 \lesssim 10^2~.
\label{eq:pre11}
\end{equation}
In addition, $\delta$ in (\ref{eq:pre5}) equals the power-law exponent
of the PDF of the observed glitch sizes,
$\tilde{X}_0 \Delta\nu_i$, 
and can therefore be estimated directly from the data.

It is tempting to reduce the search volume by marginalizing 
${\cal L}$
over one or more parameters.
The problem with doing so is that the likelihood is flatter 
(i.e.\ less informative) 
as a function of the surviving parameters than it would
otherwise be. On the other hand, one hopes that a restricted search
would rule out some of the parameter space and clarify
the broad outline of the target volume,
as a first step towards a refined, seven-parameter search.

In this paper, we experiment with two marginalization procedures.
First, we marginalize ${\cal L}$ over $X(T_1^+)$ 
subject to a uniform prior;
that is, we calculate $\int_0^1 dX(T_1^+) \, {\cal L}$
and keep the other six parameters free.
As the number of glitches rises,
${\cal L}$ depends more weakly on $X(T_1^+)$;
the memory of the initial condition fades.
Second, we exclude $\Delta\nu_1,\dots,\Delta\nu_N$
from consideration and fit $T_1,\dots,T_N$ only.
Hence we also exclude the second sum in (\ref{eq:pre7})
(involving four terms in curly braces),
so that ${\cal L}$ depends only on $\tilde{X}_0$, $\tilde{t}_0$,
$\alpha$, $X(T_1^+)$, and $\gamma$.
We find that neither marginalization procedure delivers
a substantial advantage,
so we focus on the full, seven-parameter search
in \S\ref{sec:pre4}.

\subsection{Particle swarm and Markov chain Monte Carlo algorithms
 \label{sec:pre3b}}
We conduct the full, seven-parameter maximization of ${\cal L}$
with two independent algorithms:
particle swarm (PS) optimization
and nested random sampling
(Markov chain Monte Carlo, henceforth MCMC).
The algorithms have complementary strengths.
Running both provides a useful cross-check on what is a 
challenging task.

PS is a population-based algorithm
\citep{cle10}.
A collection of `walkers' move in steps throughout a search volume.
At each step, the algorithm evaluates the objective function
at each walker and updates the walker's velocity.
We employ here the official MATLAB 
\footnote{
www.mathworks.com
}
implementation, called by the function {\tt particleswarm} in the
Global Optimization Toolbox,
e.g.\ \citet{mez11}.
The relevant algorithm control parameters are
inertia range $[0.9,2.5]$, social adjustment weight 0.35,
self adjustment weight 0.35, swarm size $6\times 10^3$,
function tolerance $10^{-7}$, and maximum stall interations $10^3$.
Other parameters are set to their default values.

MCMC is a traditional, multi-nested, random sampler
\citep{bro11}.
It evaluates the posterior directly from Bayes's Rule at a random selection
of search points. At each step, it preferentially refines the posterior,
wherever the probability density is greatest.
We employ a user-contributed MATLAB implementation
\footnote{
https://github.com/mattpitkin/matlabmultinest
} 
\footnote{
Two algorithms are offered by the MATLAB nested sampling toolbox.
Here we use the one introduced by \citet{vei10},
which replaces ellipsoidal rejection with MCMC sampling of the prior.
}
\citep{pit18}, 
which is well tested and used extensively for high-dimensional parameter estimation
by the gravitational wave data analysis community.
MCMC provides a valuable cross-check and
can be used to refine the posterior in the vicinity of a peak
discovered first by PS.
The relevant algorithm control parameters are
live number $5\times 10^3$ and tolerance $5\times 10^{-5}$,
with nested sampling activated.

\section{Parameter estimation for Poisson-like glitchers
 \label{sec:pre4}}
\begin{figure}
\centering
\includegraphics[width=1\columnwidth]{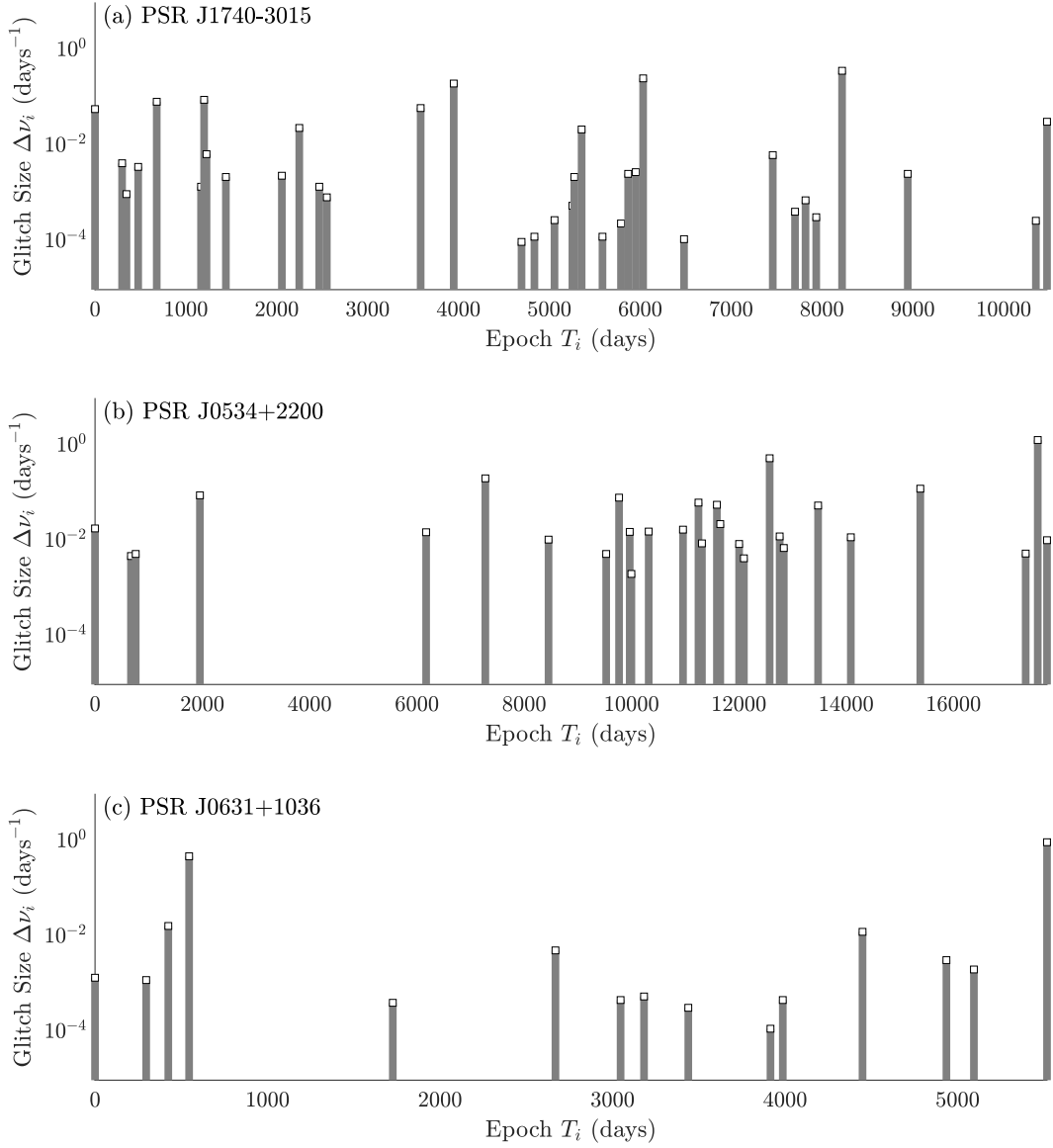}
\caption{
Time series of the observed glitch sizes $\Delta\nu_i$
(vertical axis; in ${\rm d}^{-1}$)
and epochs $T_i$
(horizontal axis; in ${\rm d}$ since the first glitch)
for PSR J1740$-$3015 (top panel),
PSR J0534$+$2200 (middle panel),
and PSR J0631$+$1036 (bottom panel).
These data are used to estimate the parameters of the 
state-dependent Poisson model in \S\ref{sec:pre4}
and predict future glitch epochs in \S\ref{sec:pre5}.
}
\label{fig:pre0}
\end{figure}

In this section, we estimate the parameters of the state-dependent
Poisson model in \S\ref{sec:pre2} using the numerical method in
\S\ref{sec:pre3} and Appendix \ref{sec:preappa}.
We focus on three objects:
PSR J1740$-$3015 ($N=35$),
PSR J0534$+$2200 ($N=28$),
and PSR J0631$+$1036 ($N=15$).
These objects are chosen, because $N$ is large enough
to make parameter estimation meaningful,
and because their waiting time and size PDFs are consistent with
exponentials and power laws respectively,
which is exactly what the state-dependent Poisson model predicts
in the $\alpha \gtrsim \alpha_{\rm c}$ regime;
see \citet{ful17} and \S\ref{sec:pre2b} above.
The observed epochs and sizes in the three objects
are plotted as time series in Figure \ref{fig:pre0}.
Other objects with relatively large samples,
e.g.\ PSR J0537$-$6910 ($N=42$)
\citep{fer18}
and PSR J0835$-$4510 ($N=21$),
glitch quasiperiodically and display approximately Gaussian size PDFs.
In its original form, the state-dependent Poisson model
in \S\ref{sec:pre2} does not apply to them,
so they are not analysed in this paper.
Likewise, PSR J1341$-$6220 ($N=23$) appears to be a 
quasiperiodic-Poisson hybrid
\citep{how18,fue19}
and also falls outside the scope of the analysis below.

\subsection{Corner plots
 \label{sec:pre4a}}
\begin{figure}
\centering
\includegraphics[width=1\columnwidth]{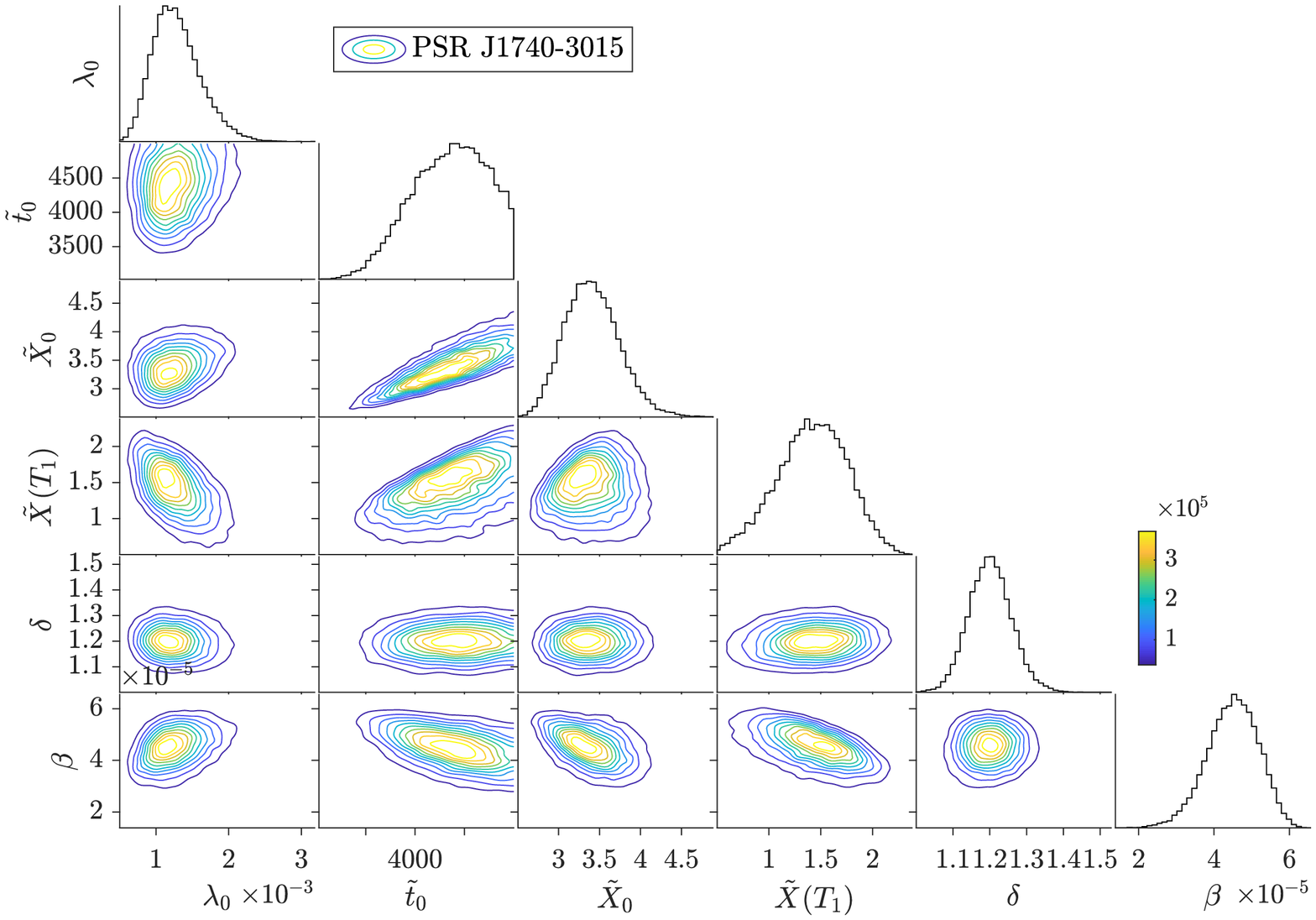}
\caption{
Corner plot of the likelihood for PSR J1740$-$3015
marginalized over all \hl{$^6 C_2 = 15$} combinations of the parameters 
\hl{$\theta=\{ \lambda_0, \tilde{t}_0, \tilde{X}_0, \tilde{X}(T_1^+), \delta, \beta \}$}.
Ten evenly spaced contours are drawn between the maximum and minimum likelihoods 
in the MCMC sample, color-coded according to the color bar.
Units: $\lambda_0$ in ${\rm d}^{-1}$, $\tilde{t}_0$ in {\rm d},
$\tilde{X}_0$ and $\tilde{X}(T_1^+)$ in ${\rm rad \, d^{-1}}$.
}
\label{fig:pre1}
\end{figure}

\begin{figure}
\centering
\includegraphics[width=1\columnwidth]{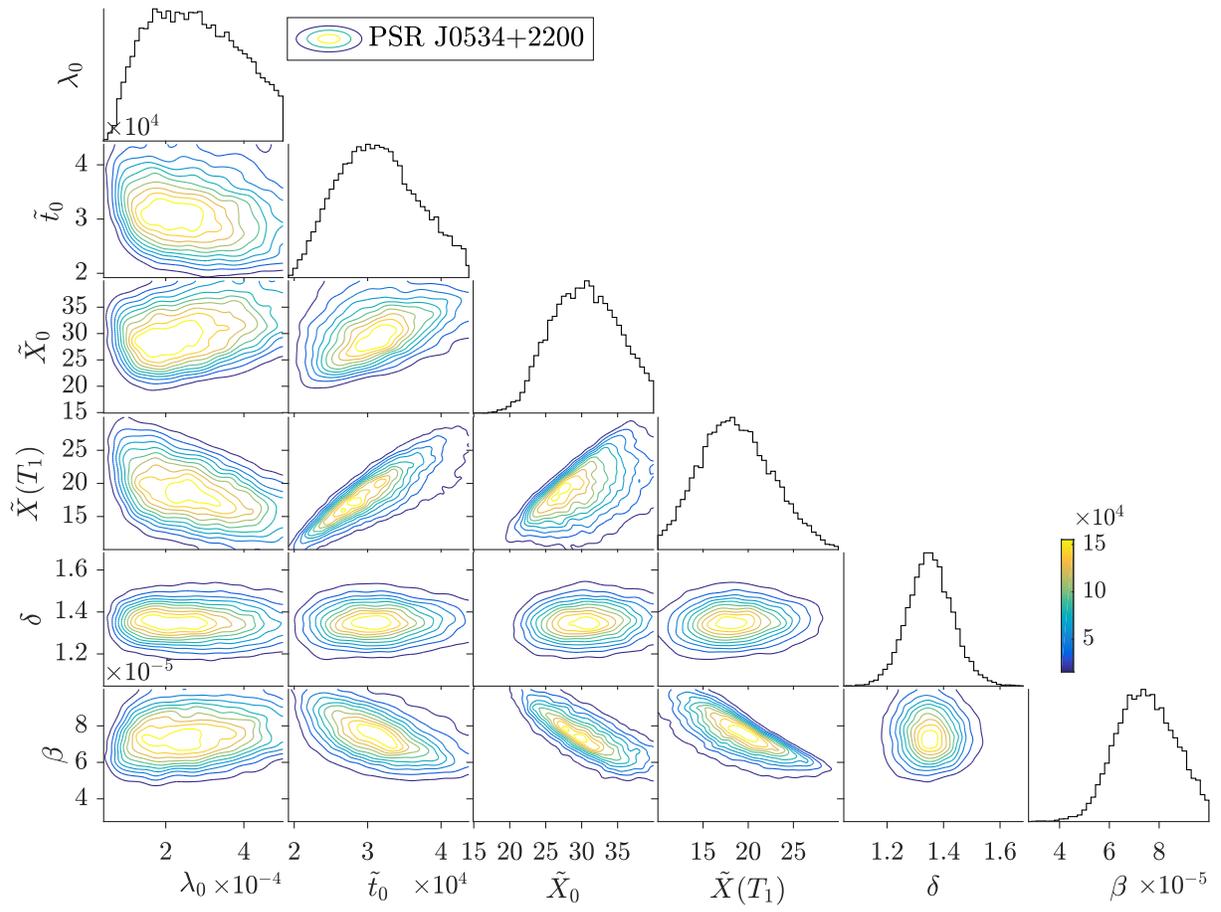}
\caption{
As for Figure \ref{fig:pre1} but for PSR J0534$+$2200.
}
\label{fig:pre2}
\end{figure}

\begin{figure}
\centering
\includegraphics[width=1\columnwidth]{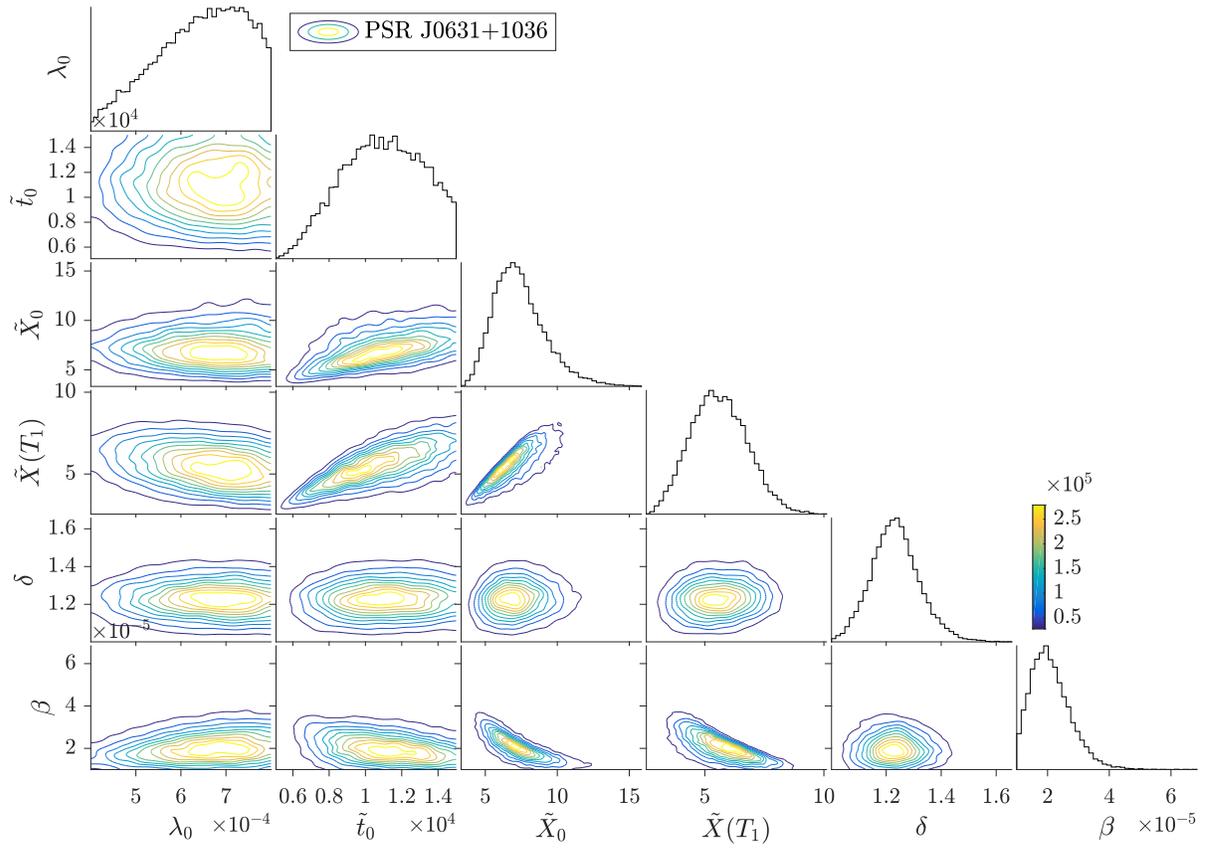}
\caption{
As for Figure \ref{fig:pre1} but for PSR J0631$+$1036.
}
\label{fig:pre3}
\end{figure}

Figures \ref{fig:pre1}, \ref{fig:pre2}, and \ref{fig:pre3} 
summarize the results of the MCMC estimation exercise pictorially for 
PSR J1740$-$3015, PSR J0534$+$2200, and PSR J0631$+$1036 respectively.
Numerical values for the MCMC and PS maximum likelihood estimates of
$\lambda_0$, $\tilde{t}_0$, $\tilde{X}_0$, $\tilde{X}(T_1^+)$, $\delta$, and $\beta$
are quoted in Table \ref{tab:pre1} for the three objects.
We start with $\gamma=2\pi$ for numerical convenience and extend to
a range of $\gamma$ values in \S\ref{sec:pre4c}.
Figures \ref{fig:pre1}--\ref{fig:pre3} 
are traditional corner plots.
Every panel featuring colored contours corresponds to the likelihood
marginalized over all but two parameters
(e.g.\ $\lambda_0$ and $\beta$ in the bottom-left corner);
yellow (blue) contours correspond to 
$\approx 0.9$ (0.09) times the maximum.
\footnote{
Strictly speaking, the functions plotted in Figures \ref{fig:pre1}--\ref{fig:pre3} 
are marginalized likelihoods rather than posterior PDFs.
The MCMC sampler explores the shape of ${\cal L}$ around the peak,
but no attempt is made to calculate Bayesian evidences,
and the priors are assumed to be uninformative
inside the definitional bounds in \S\ref{sec:pre3a}.
}
Every panel featuring a single, black curve corresponds to the likelihood
marginalized over all parameters but one
(e.g.\ $\lambda_0$ in the top-left corner).
In Table \ref{tab:pre1},
the PS maximum likelihood estimates are quoted without error intervals,
because we do not have the computational resources at our disposal
to run a systematic suite of PS searches with different initializations
for a seven-dimensional problem.
However, the MCMC maximum likelihood estimates come with an
error interval automatically attached 
(one-sigma half-width of the likelihood function).

\begin{table}
\begin{center}
\begin{tabular}{llccccccc} \hline
PSR J & Method & $\lambda_0$ & $\tilde{t}_0$ & 
 $\tilde{X}_0$ & $\tilde{X}(T_1^+)$ & $\delta$ & $\beta$ \\ \hline
1740$-$3015 & MCMC & $1.29\times10^{-3}$ & $4.35\times10^3$ & 
 3.42 & 1.41 & 1.20 & $4.5\times10^{-5}$ \\
& & $\pm 3.3\times10^{-4}$ & $\pm 3.8\times10^2$ & 
 $\pm 0.34$ & $\pm 0.34$ & $\pm 0.06$ & $\pm 7\times10^{-6}$ \\
& PS & $1.04\times10^{-3}$ & $3.78\times10^3$ & 
 2.89 & 1.34 & 1.21 & $5.5\times10^{-5}$ \\
0534$+$2200 & MCMC & $2.72\times10^{-4}$ & $3.14\times10^4$ & 
 30.5 & 18.7 & 1.35 & $7.5\times10^{-5}$ \\
& & $\pm 1.1\times10^{-4}$ & $\pm 5.3\times10^3$ & 
 $\pm 4.5$ & $\pm 3.9$ & $\pm 0.08$ & $\pm 1\times10^{-5}$& \\
& PS & $1.08\times10^{-4}$ & $2.90\times10^{4}$ & 
 25.9 & 18.5 & 1.36 & $8.5\times10^{-5}$ & \\
0631$+$1036 & MCMC & $6.40\times10^{-4}$ & $1.10\times10^4$ & 
 7.59 & 5.74 & 1.23 & $2.0\times10^{-5}$ \\
& & $\pm 9.7\times10^{-5}$ & $\pm 2.2\times10^3$ & 
 $\pm 1.8$ & $\pm 1.2$ & $\pm 0.09$ & $\pm 6\times10^{-6}$ \\
& PS & $4.91\times10^{-4}$ & $1.04\times10^4$ & 
 6.70 & 6.04 & 1.24 & $2.3\times10^{-5}$ \\
\hline
\end{tabular}
\end{center}
\caption{
MCMC and PS maximum likelihood estimates of the six parameters
of the state-dependent Poisson process in \S\ref{sec:pre2} with $\gamma=2\pi$.
The first (second) row of each MCMC entry contains the mean (standard deviation)
of the likelihood function.
Units: $\lambda_0$ in ${\rm d}^{-1}$, $\tilde{t}_0$ in {\rm d},
$\tilde{X}_0$ and $\tilde{X}(T_1^+)$ in ${\rm rad \, d^{-1}}$.
}
\label{tab:pre1}
\end{table}

Despite the lack of formal confidence intervals on the PS estimates,
we conclude that the PS and MCMC results are consistent on the whole.
The PS estimate lies within the two-sigma MCMC error bar for every
entry in Table \ref{tab:pre1} and within the one-sigma MCMC error bar 
for every entry except $\lambda_0$ and $\tilde{X}_0$ in PSR J0534$+$2200, 
$\tilde{t}_0$, $\tilde{X}_0$, and $\beta$ in PSR J1740$-$3015,
and $\lambda_0$ in PSR J0631$+$1036.
All the one- and two-variable likelihoods are unimodal.
A detailed sensitivity analysis is computationally expensive and
lies outside the scope of this paper.
Instead we discuss what the estimates teach us physically,
and how reasonable they are from the physical perspective,
in \S\ref{sec:pre4b} and \S\ref{sec:pre4c}.

\subsection{Rate variables: $\tilde{t}_0$ and $\lambda_0$
 \label{sec:pre4b}}
The characteristic time-scale $\tilde{t}_0$ is $\sim 10$ times
the mean observed waiting time for all three objects in Table \ref{tab:pre1}.
This is reasonable;
the waiting times are bounded above by $\tilde{t}_0$
in the model in \S\ref{sec:pre2}.

The microscopic trigger rate $\lambda_0$ is inferred to equal
$\approx 6\tilde{t}_0^{-1}$ 
for all three objects in Table \ref{tab:pre1}.
In other words, $\lambda_0$ is within a factor of $\approx 2$ 
of the mean observed glitch rate.
Again this is reasonable,
because $X(t)$ spends most of the time fluctuating around $X(t)\sim 0.5$,
yielding $\lambda[X(t)] \sim \lambda_0$.
Interestingly, the maximum likelihood estimates yield
$\alpha = \lambda_0 \tilde{t}_0 \approx 6$,
cf.\ $\alpha_{\rm c}(\beta) \sim 1$.
In other words, the objects spin down under but close to the critical rate
and exhibit power-law size and exponential waiting time PDFs as expected.

\subsection{Size variables: $\tilde{X}_0$, $\gamma$, $\beta$, and $\delta$
 \label{sec:pre4c}}
The characteristic angular velocity scale, $\tilde{X}_0$,
ranges between $5\,\mu{\rm Hz}$ and $56\,\mu{\rm Hz}$
for the three objects in Table \ref{tab:pre1}.
From the observational perspective, this is reasonable.
The above numbers are $\approx 10$ times larger than the maximum
glitch size observed,
and glitch size is bounded above by $\tilde{X}_0$
in the model in \S\ref{sec:pre2}.
From the theoretical perspective,
$\tilde{X}_0 = X_{\rm c}$ is consistent with sensible values 
of the maximum critical angular velocity lag 
in the superfluid vortex avalanche picture
\citep{lin91,war11},
viz.\
\begin{equation}
 X_{\rm c}
 \leq
 6\times 10^{-2}
 \left(
  \frac{F_{\rm max}}{{\rm kev\,fm^{-1}}}
 \right)
 \left(
  \frac{\rho}{10^{13}\,{\rm g\,cm^{-3}}}
 \right)^{-1}
 \left(
  \frac{l}{10^{2}\,{\rm fm}}
 \right)^{-1}
 \, {\rm Hz}~,
\label{eq:pre12}
\end{equation}
where $F_{\rm max}$ is the maximum pinning force per nuclear lattice site,
$\rho$ is the superfluid mass density,
and $l$ is the pinning site separation.
An analogous expression for the critical crustal stress in the starquake picture
can be deduced from the models proposed by
\citet{mid06} and \citet{akb18};
see also \S{5} in \citet{mel18}.
The observed lack of strong size-waiting-time correlations
in Poisson-like glitchers implies that the true critical lag is small
compared to the maximum critical lag in (\ref{eq:pre12});
the stress reservoir never empties completely.

The minimum fractional avalanche size, $\beta$,
satisfies $\beta \lesssim 10^{-4} \ll 1$ 
for the three objects in Table \ref{tab:pre1}.
In a fitting exercise of this kind, $\beta$ is expected to reflect
the dynamic range of the observed glitch sizes.
The observed glitches span $\approx 4\,{\rm dex}$ in the three objects,
consistent with the maximum likelihood estimates of $\beta$.
If future observations reveal even smaller events,
the $\beta$ estimate will decrease.
Conversely, if the smallest glitches observed to date are set by
the resolution of the timing experiment,
then the estimated value of $\beta$ is higher than the true,
underlying, physical value,
e.g.\ \citet{jan06} but cf. \citet{esp14} for PSR J0534$+$2200.

The power-law exponents of the avalanche size PDF and observed
glitch size PDF are equal in the theory in \S\ref{sec:pre2}.
Hence the PS and MCMC estimates of $\delta$ agree closely;
they both reflect the well-defined shape of the observed glitch size PDF.
They are also in accord with previous estimates of this quantity
\citep{mel08,ash17,how18,sha18}.

It turns out that maximizing ${\cal L}$ over $\gamma$ as well as the
six parameters in Table \ref{tab:pre1} is too unwieldy even with
human supervision, given the computational resources at our disposal.
We therefore repeat the maximization in Table \ref{tab:pre1} for
$\gamma=10$, 20, and 30 to gain an idea of how the results depend
on $\gamma$.
Overall, the dependence is weak:
(i)
$\lambda_0$ and $\tilde{t}_0$ change by $\lesssim 50$ per cent
in opposite senses across the range, 
limiting the variation in $\alpha=\lambda_0 \tilde{t}_0$ to $\lesssim 10$ per cent;
(ii)
$\tilde{X}_0$ and $\tilde{X}(T_1^+)$ increase in rough proportion to $\gamma$,
consistent with the appearance of $\gamma$ in (\ref{eq:pre7})
through the product $\gamma\Delta\nu_i$;
(iii) 
$\beta$ decreases in rough proportion to $\gamma$;
and
(iv) 
$\delta$ changes by $\lesssim 10$ per cent.

When more data accumulate in the future,
it will be interesting to study $\gamma$ in greater detail.
The ratio of the crust and superfluid moments of inertia,
which sets $\gamma$, is the subject of considerable theoretical uncertainty
\citep{lin99,and12,cha13,pie14,eya17}.
If the glitch-related angular momentum reservoir resides in the core superfluid,
one expects $I_{\rm s}/I_{\rm c} \sim 10^2$ and $\gamma \lesssim 10^3$.
Alternatively, if the reservoir resides in the inner crust,
and the rigid crust and core superfluid are coupled magnetically
by interactions between superfluid vortices and superconductor flux tubes,
then one expects $I_{\rm s}/I_{\rm c} \sim 10^{-2}$
and $\gamma\approx 1$.
The exact value of $\gamma$ depends on subtle microphysics,
e.g.\ the strength of entrainment
\citep{cha13}
and the degree to which the vortices and flux tubes are tangled
\citep{dru18}.
Larger data sets have the potential to shed light on these interesting issues.

\subsection{Stress history
 \label{sec:pre4d}}
\begin{figure}
\centering
\includegraphics[width=0.8\columnwidth]{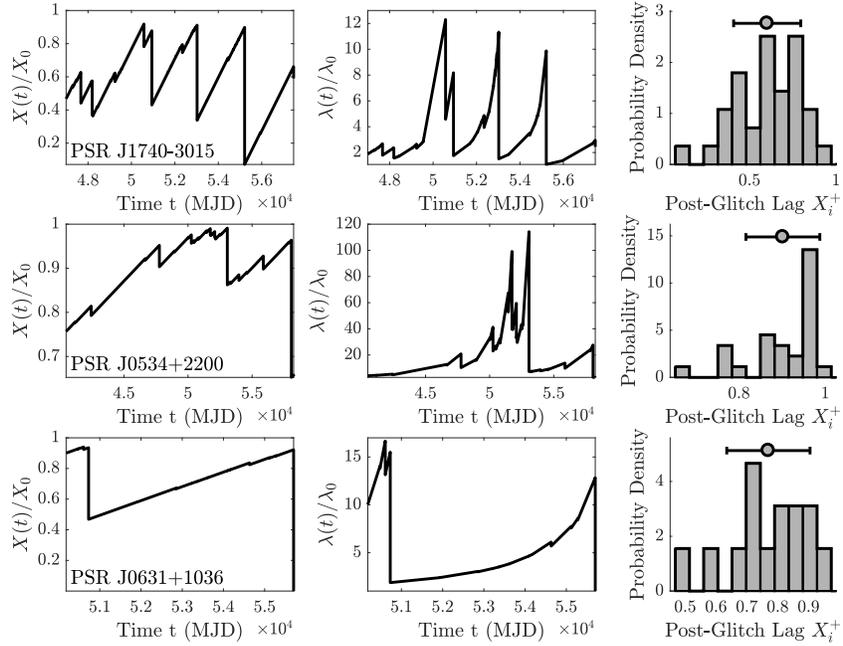}
\caption{
Evolution of the stress variable in 
PSR J1740$-$3015 (top row),
PSR J0534$+$2200 (middle row),
and PSR J0631$+$1036 (bottom row). 
{\em Left column.}
Maximum likelihood stress history, $X(t)$,
normalized by the critical stress.
{\em Middle column.}
Dimensionless instantaneous glitch rate,
$\lambda[X(t)]/\lambda_0=[1-X(t)]^{-1}$,
versus time, $t$ (in MJD).
{\em Right column.}
Histogram of the normalized post-glitch stress,
$X(T_i^+)$ ($1\leq i \leq N$).
}
\label{fig:pre4}
\end{figure}

Figure \ref{fig:pre4} displays the maximum likelihood estimates
of the stress histories of the three objects in Table \ref{tab:pre1}.
We point out some key features.
First, the trigger rate $\lambda(X)$ (middle column)
varies substantially in all three objects,
e.g.\ by a factor of \hl{$\approx 11$} from trough to peak in PSR J1740$-$3015.
This is a general and slightly counterintuitive feature of the model
in \S\ref{sec:pre2}:
the rate is variable, even though the waiting time PDF is consistent
with a constant-$\lambda$ exponential to a good approximation;
see \S{8} and Figure 24 in \citet{ful17}.
Second, $X(t)$ (left column) zig-zags more violently in PSR J1740$-$3015
than in PSR J0534$+$2200 and PSR J0631$+$1036.
This is consistent with the $\delta$ estimates
in the following sense.
The long-term average $\langle \Delta \nu_i \rangle$ 
is dominated by the largest --- and rarest --- events for 
$\eta[X(T_i^-) - \gamma\Delta\nu_i | X(T_i^-)] 
 \propto (\Delta \nu_i)^{-\delta}$
and $1 < \delta < 2$.
Hence it is normal for $X(t)$ to increase steadily over a 
relatively long interval,
its rise interrupted by minor avalanches,
before a large event strikes and resets $X(t)$ nearly to zero.
Large events occur most frequently in PSR J1740$-$3015,
whose $\delta$ value is the smallest of the three.
Third, the histogram of the stresses $X(T_i^+)$ ($1\leq i \leq N$)
immediately after each glitch (right column) 
peaks at $0.5 \lesssim X(T_i^+) \lesssim 1$,
consistent with the prediction by \citet{ful17};
see Figure 21 in the latter reference.

\begin{figure}
\hspace{-1.5cm}
\includegraphics[width=1.2\columnwidth]{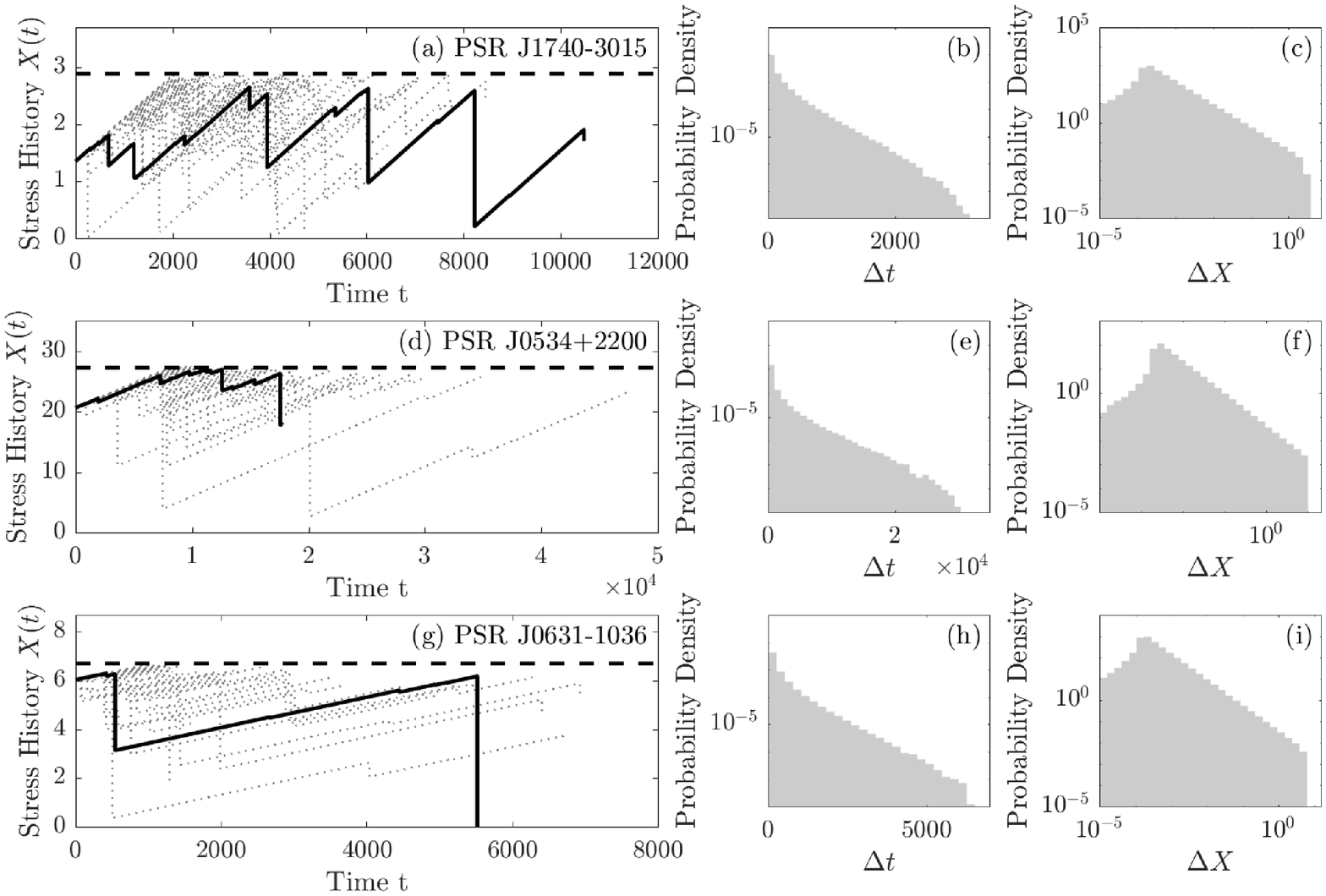}
\caption{
Stress history $X(t)$ (left column),
waiting time PDF $p(T_{i+1}-T_i)$ (middle column; log linear axes), 
and size PDF $p(\gamma\Delta \nu_i)$ (right column; log log axes)
from Monte Carlo simulations of the model in \S\ref{sec:pre2}
with the maximum likelihood estimate of $\theta$.
The black, solid curves and grey, dotted curves indicate
best-fitting and alternative stress histories respectively.
Top row: PSR J1740$-$3015.
Middle row: PSR J0534$+$2200.
Bottom row: PSR J0631$+$1036.
Parameters: as in Table \ref{tab:pre1}; $10^6$ realizations per object.
}
\label{fig:pre5}
\end{figure}

In Figure \ref{fig:pre5}, we present an ensemble of stress histories $X(t)$
(left column)
generated by Monte Carlo iteration of (\ref{eq:pre1})--(\ref{eq:pre5}).
For each object in Table \ref{tab:pre1},
equations (\ref{eq:pre1})--(\ref{eq:pre5}) are evaluated with the
maximum likelihood estimate of the parameter vector $\theta$.
It is clear by eye that many of the Monte Carlo stress histories (grey curves)
resemble qualitatively the one that fits the observational data best 
(maximum ${\cal L}$; black curve).
The size and waiting time cumulative distribution functions (CDFs)
in the middle and right columns are also consistent with the data.
Figure \ref{fig:pre5} therefore serves as a useful cross-check,
that the maximum likelihood fit is not a statistical outlier.

The log likelihood (\ref{eq:pre7}) does not adjust for the
measurement uncertainties in $T_i$ and $\Delta\nu_i$,
because the formal errors quoted in the discovery papers are small
(typically $\lesssim 5$ per cent in both variables),
and the systematic errors are hard to quantify
\citep{jan06,mel08,esp11,yu17}.
Generalizing ${\cal L}$ to include measurement uncertainties
is a topic for future work.
One interesting question is whether or not larger glitches,
whose fractional measurement uncertainties are typically lower,
are more informative about the maximum likelihood stress history
than smaller glitches.
In the analysis presented in this paper, the answer is no:
all observed glitches above the minimum size set by $\beta$
in (\ref{eq:pre5}) are treated on an equal footing.
In a fuller analysis including measurement uncertainties,
the answer may be yes.
Irrespective of measurement uncertainties, however,
the constraint $\gamma\Delta\nu_i \leq X(T_i^-)$
leading to (\ref{eq:pre10}) implies that larger glitches probe intervals,
when $X(t)$ is closer to $X_{\rm c}$,
and may influence the maximum likelihood estimate
of $\tilde{X}_0$ more than smaller glitches. 
One can also ask: are larger glitches predicted more reliably
than smaller glitches by the maximum likelihood model?
Again the answer seems to be no.
When we generate multiple Monte Carlo stress histories using
the maximum likelihood estimate of $\theta$, as in Figure \ref{fig:pre5},
the simulation is no better at predicting larger glitches at the
observed epochs than smaller glitches,
because the sizes are distributed as a power law and
depend weakly on $X(T_i^-)$ much of the time.
Likewise, as the dispersion in waiting times for a Poisson process
equals the mean,
there is no reason to expect the Monte Carlo epochs
in Figure \ref{fig:pre5} to coincide closely with
the observed epochs on an individual basis.
We merely expect the state-dependent Poisson model to predict
epochs better than a homogeneous Poisson model on average,
as discussed further in \S\ref{sec:pre5}.

\section{Epoch prediction
 \label{sec:pre5}}
Even though the trigger rate $\lambda(X)$ varies $\gtrsim 10$-fold 
over the time intervals that PSR J1740$-$3015, PSR J0534$+$2200,
and PSR J0631$+$1036 have been observed,
the waiting time PDFs in all three objects are approximately exponential,
as one would obtain for a homogeneous (i.e.\ constant-$\lambda$) Poisson process.
It is therefore important to ask:
is the predictive power of the state-dependent Poisson process in \S\ref{sec:pre2}
any greater than a homogeneous Poisson process with the same average rate?
The answer should be yes,
because the model in \S\ref{sec:pre2} incorporates the extra information
contained in the glitch sizes via the statistical correlation between
$T_i - T_{i-1}$ and $X(T_i^-)$ in (\ref{eq:pre2}).
We quantify the relative performance below.

\subsection{Validation against $T_2,\dots,T_N$
 \label{sec:pre5a}}
\begin{figure}
\centering
\includegraphics[width=0.48\columnwidth]{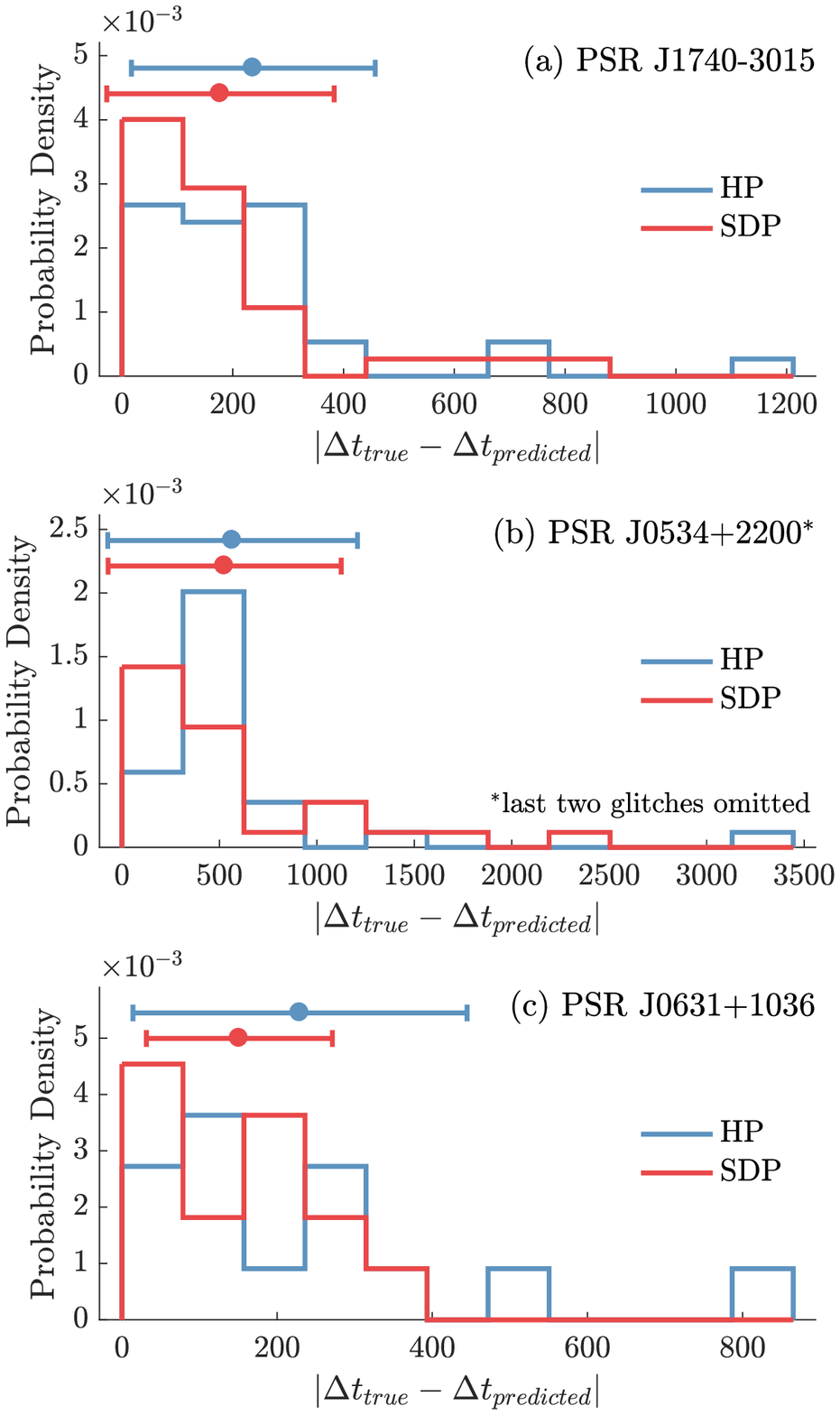}
\caption{
Histograms of the unsigned absolute error in the predicted epoch, 
i.e.\ the predicted minus the true epoch (measured in days),
for $T_2,\dots,T_N$ in 
(a) PSR J1740$-$3015, (b) PSR J0534$+$2200 and (c) PSR J0631$+$1036
for the state-dependent (red border)
and homogeneous (blue border) Poisson models.
The red and blue histograms overlap in some bins.
Color-coded dots and horizontal error bars indicate the mean
and standard deviation respectively for both models;
see also Table \ref{tab:pre2}.
}
\label{fig:pre6}
\end{figure}

We start by analysing the statistics of the errors in the epoch predictions
of the state-dependent and homogeneous Poisson models when back-tested 
{\em post factum} against the historical epochs $T_2,\dots,T_N$.
Figure \ref{fig:pre6}
displays histograms of the unsigned absolute error
(i.e.\ absolute value of the predicted minus the true epoch)
for all $T_i$ with $2\leq i \leq N$
for the three objects in Table \ref{tab:pre1}.
\footnote{
For example the state-dependent Poisson model predicts
$T_N = 57654\pm 251$, $60814\pm 2104$, and $55570\pm 257$
(all in MJD; one-sigma uncertainties)
for PSR J1740$-$3015, PSR J0534$+$2200, and PSR J0631$+$1036 respectively
using the data from events $2\leq i \leq N-1$.
The observed epochs,
$T_N = 57469$, $58237$, and $55702$ respectively,
lie within two sigma of the predictions.
}
The associated means and standard deviations are summarized in Table \ref{tab:pre2}.
We find, as expected, that the state-dependent Poisson process
is a better predictor than the homogeneous Poisson process.
Importantly, though, the advantage is modest for the small samples
available at present
and it can be vitiated by an abnormally large, recent glitch.
The mean error produced by the state-dependent Poisson model is
$77$, $98$, and $73$ per cent of that produced by the
homogeneous Poisson model 
for PSR J1740$-$3015, PSR J0534$+$2200 (minus the latest two glitches), 
and PSR J0631$+$1036 respectively.
However, if the large, penultimate glitch in PSR J0534$+$2200 
and its successor are included,
the mean error of the state-dependent Poisson model rises to $108$ per cent
of the homogeneous Poisson model (second line of Table \ref{tab:pre2}).
It drops back to $98$ per cent,
if the first four glitches in PSR J0534$+$2200 are excluded
(fourth line of Table \ref{tab:pre2}),
from the time before daily monitoring of the object commenced,
when some small events may have been missed.
For all three objects, the difference between the mean errors of the two models
is smaller than their standard deviations.
The dispersion in waiting times for any sort of Poisson process
is typically comparable to the mean,
due to the exponential form of (\ref{eq:pre2}),
so wide error bars are the rule.
In summary, therefore, the state-dependent Poisson model does not enjoy
an unqualified advantage in epoch prediction over the homogeneous Poisson model
for samples with $N\leq 35$ like those in Figure \ref{fig:pre6}.

\begin{table}
\begin{center}
\begin{tabular}{lcccc}\hline
 PSR J & Mean (SDP) & Std dev (SDP) & Mean (HP) & Std dev (HP) \\ \hline
 1740$-$3015 & 183.4 & 205.9 & 237.2 & 220.1 \\
  0534$+$2200 & 592.8 &  682.4 & 551.2 & 654.6 \\
 0534$+$2200\tablenotemark{\ast} & 563.5 &  607.1 & 573.7 & 666.5 \\
 0534$+$2200\tablenotemark{\ast\ast} & 381.6 &  309.0 & 389.5 & 303.7 \\
 0631$+$1036 & 167.2 & 112.7 & 229.6 & 215.3 \\
 \hline
\end{tabular}
\end{center}
\caption{
Means and standard deviations of the error histograms 
in Figure \ref{fig:pre6}
for the state-dependent (SDP) and homogeneous (HP) Poisson models. 
All entries are measured in days.
}
\tablenotetext{\ast}{Excludes the large, penultimate glitch and its successor
 ($27\leq i \leq 28$).}
\tablenotetext{\ast\ast}{Excludes the glitches before daily monitoring commenced
 ($1\leq i \leq 4$).}
\label{tab:pre2}
\end{table}

\begin{figure}
\centering
\includegraphics[width=0.5\columnwidth]{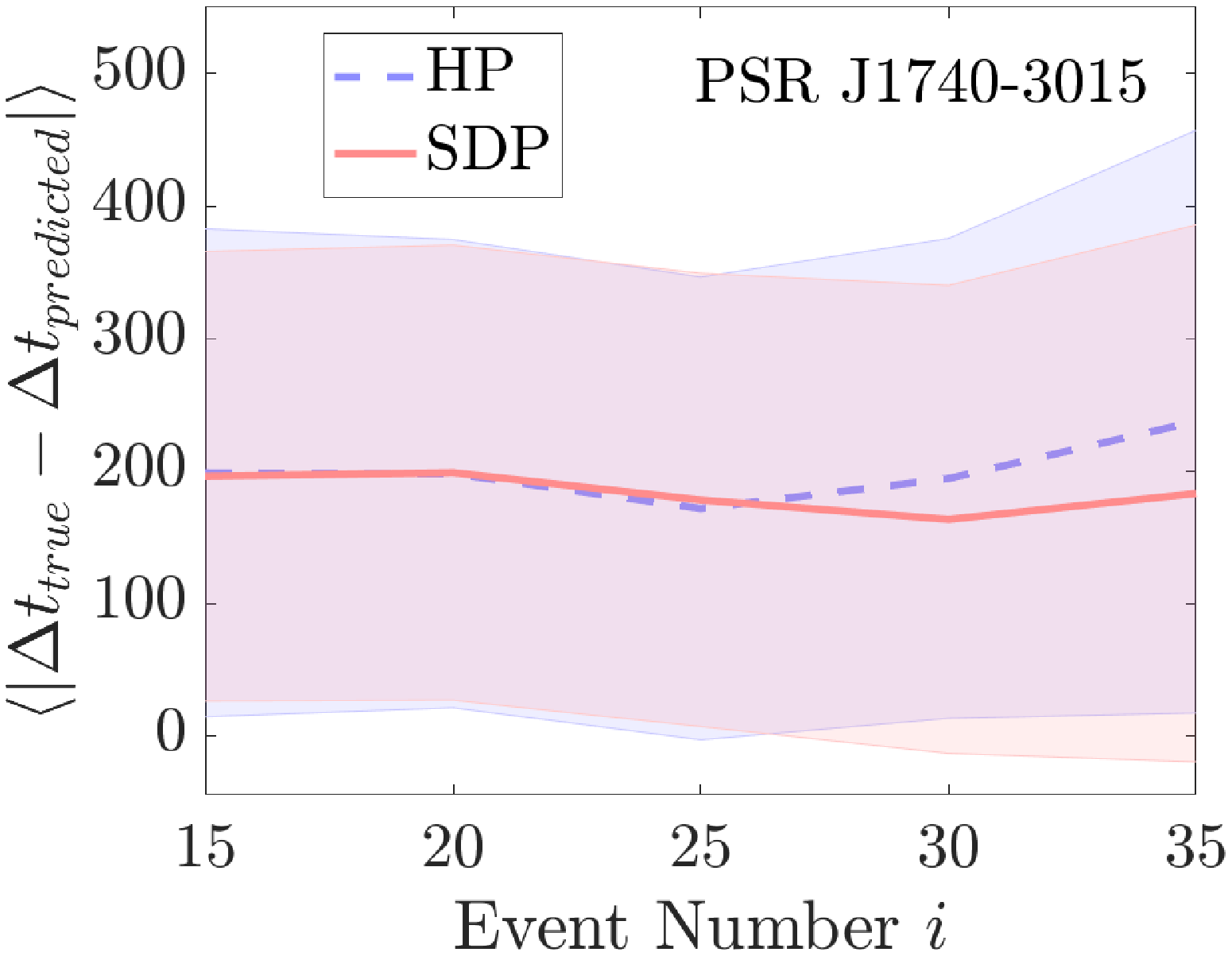}
\caption{
Unsigned absolute error in the predicted epoch (in days)
versus event number $i$ ($15\leq i \leq 35$) for the state-dependent
(solid red curve) and homogeneous (dashed blue curve) Poisson models
for PSR J1740$-$3015. 
The shading (with the same color coding) denotes the one-sigma error bars for the
two models.
Events with $2\leq i \leq 14$ are not plotted,
because the scatter is large and uninformative.
}
\label{fig:pre9}
\end{figure}

The histograms in Figure \ref{fig:pre6}
unavoidably blend together two factors:
the goodness-of-fit of each model,
and the improved ability of each model to make predictions,
as the sample of events grows.
For example, both models are expected to perform better at predicting
$T_{100}$ than $T_2$, statistically speaking.
This is illustrated in Figure \ref{fig:pre9},
where the error is graphed versus event number for both Poisson models. 
PSR J1740$-$3015 is chosen for this figure 
because it has the largest number of glitches.
The early events ($2\leq i \leq 14$) are not plotted,
because the scatter is large and uninformative;
both models struggle to predict epochs with insufficient information.
For intermediate events ($15\leq i \lesssim 25$),
the scatter is lower,
and the errors produced by the two models are comparable. 
For later events ($25 \lesssim i \leq 35$), 
the red and blue curves separate,
and the state-dependent Poisson model performs slightly better.
Again, though, the performance difference is small compared to the dispersion.
There is no strong, sustained decrease in the error produced by
the state-dependent Poisson model over the range $15\leq i \leq 35$,
confirming that the existing event sample is small.

Ideally, the histograms in Figure \ref{fig:pre6}
would be redrawn for (say) $T_{50}$ for a large number of objects,
but this is impossible with existing data.
Instead we perform a numerical Monte Carlo simulation along 
the same lines and present the results in Appendix \ref{sec:preappb}.
It is found that the predictive advantage of the state-dependent Poisson model
asserts itself increasingly for $N\gtrsim 50$,
as the information embedded in the glitch sizes
increasingly constrains $X(t)$ and hence $\lambda[X(t)]$.
That is, the mean error from the state-dependent Poisson model
is lower than from the homogeneous Poisson model,
and the standard deviation scales $\propto N^{-1/2}$ as expected.
The Monte Carlo trend in the standard deviation contrasts somewhat with 
the real data from PSR J1740$-$3015,
analysed in Figure \ref{fig:pre9},
where there is no obvious shrinkage in the shaded red band 
at $i\approx 35$ relative to $i\lesssim 25$.
It is unclear whether or not this is a statistical accident;
more data are needed to resolve the issue.
We emphasize that the results in Appendix \ref{sec:preappb}
apply in an astrophysical setting only to the extent that the glitch microphysics 
conforms to the meta-model of a state-dependent Poisson process.

\subsection{The next glitch: $T_{N+1}$
 \label{sec:pre5b}}
We conclude this section by presenting falsifiable epoch predictions 
for the next glitches to occur in the three objects in Table \ref{tab:pre1}.
Epochs and one-sigma error bars are tabulated in Table \ref{tab:pre3}
for the state-dependent Poisson model in \S\ref{sec:pre2}.
Assuming the model is a faithful description
of Poisson-like glitch activity,
the central tendencies of the predicted epochs are
$T_{36}=57784$ (MJD) for PSR J1740$-$3015,
$T_{29}=60713$ (MJD) for PSR J0534$+$2200,
and $T_{16}=57406$ (MJD) for PSR J0631$+$1036.
Analogous, falsifiable predictions can be made for these or other
Poisson-like glitchers in the future following the recipe in this paper,
as more events are discovered.

\begin{table}
\begin{center}
\begin{tabular}{lcc} \hline
 PSR J & Mean $T_{N+1}$ & Std dev \\ \hline
 1740$-$3015 & 57784 & 256.8 \\
 0534$+$2200 & 60713 & 1935 \\
 0631$+$1036 & 57406 & 1444 \\ \hline
\end{tabular}
\end{center}
\caption{
Mean predicted epoch (MJD) and one-sigma uncertainty
(in days)
of the next glitch in each of the objects in Table \ref{tab:pre1},
based on the state-dependent Poisson model.
}
\label{tab:pre3}
\end{table}

\section{Conclusion
 \label{sec:pre6}}
A state-dependent Poisson process of the form (\ref{eq:pre1})--(\ref{eq:pre5})
generates power-law size and exponential waiting time PDFs
in the slow-spin-down regime $\alpha \gtrsim \alpha_{\rm c} \sim 1$.
It is a plausible meta-model for glitch activity in pulsars
that do not glitch quasiperiodically,
whether the glitch microphysics involves starquakes, 
superfluid vortex avalanches, or some other stress growth-release mechanism.
In this paper we compute maximum likelihood estimates for the
seven parameters governing the dynamics described by 
(\ref{eq:pre1})--(\ref{eq:pre5})
for the three Poisson-like pulsars with the most events:
PSR J1740$-$3015, PSR J0534$+$2200, and PSR J0631$+$1036.
We find that MCMC and PS maximization algorithms produce similar estimates, with 
$3 \leq \alpha \leq 9$,
trigger rate
$1\times 10^{-4} \leq \lambda_0/(1\,{\rm d^{-1}}) \leq 1\times 10^{-3}$,
stress threshold
$3 \leq X_{\rm c} / (1\,{\rm rad \, d^{-1}}) \leq 31$,
minimum fractional avalanche size
$2\times 10^{-5} \leq \beta \leq 9\times 10^{-5}$,
and a weak dependence on the moment-of-inertia ratio
($\alpha \approx {\rm constant}$,
 $X_c \propto \gamma$, $\beta \propto \gamma^{-1}$ approximately)
in the fiducial range $6\leq \gamma \leq 30$.
The maximum likelihood estimates are tabulated in Table \ref{tab:pre1}
and cross-checked for consistency against Monte Carlo simulations.
Parameter estimation of this sort is a promising tool for probing
the glitch microphysics in the future,
when more data become available.
For example, improved estimates of $\lambda_0$ and $\gamma$
may shed light on the glitch trigger mechanism and site of the
angular momentum reservoir respectively.

The trigger rate $\lambda[X(t)] =\lambda_0[1-X(t)]^{-1}$ varies
more than 10-fold in the inferred maximum likelihood stress histories
of the above three objects,
so it is pertinent to ask whether or not a state-dependent Poisson model
delivers more accurate epoch predictions than a homogeneous Poisson model.
Back-testing against historical glitches indicates that the answer is
a qualified yes for the above three objects,
but the improvement is small both fractionally (7--27 per cent)
and compared to the dispersion and can be reversed by an abnormally large,
recent glitch (e.g.\ in PSR J0534$+$2200).
Monte Carlo simulations indicate that a sample with $N\gtrsim 50$ is needed,
in order to test reliably whether or not
the state-dependent Poisson model outperforms.
Falsifiable predictions are made that the epoch of the next glitch
in the three tested objects is
$T_{36}=57784$ (MJD) for PSR J1740$-$3015,
$T_{29}=60713$ (MJD) for PSR J0534$+$2200,
and $T_{16}=57406$ (MJD) for PSR J0631$+$1036,
with associated one-sigma error intervals.
The predicted epochs stand alongside falsifiable predictions
concerning auto- and cross-correlations between sizes and waiting times, 
which can be tested independently
\citep{mel18,car19c}.
For example, \citet{mel18} identified PSR J0205$+$6449
as one of five pulsars likely to exhibit a strong cross-correlation
between sizes and forward waiting times with the advent of more data
[see Table 2 in \citet{mel18}].
The prediction found support recently in a reanalysis 
based on twice as many events 
\citep{fue19}
but it could have been falsified just as easily.
We emphasize again that the state-dependent Poisson model
does not apply to quasiperiodic glitchers in the form introduced
by \citet{ful17}.

In order to make the most of the opportunity for falsification,
more glitches need to be discovered.
As well as monitoring more objects with higher duty cycle,
e.g.\ with phased arrays \citep{kra10,cal16},
it is worth exploring how to reduce the minimum resolvable glitch size,
in case there is a populous tail of smaller glitches waiting to be discovered
in some objects
[although arguably not in PSR J0534$+$2200; see \citet{esp14}].
Algorithms that harness the power of distributed volunteer computing
\citep{cla17}
or take a Bayesian approach to inferring glitch parameters
\citep{sha16}
are set to play a role in delivering these and other improvements.

\acknowledgments
This research was supported by the Australian Research Council
Centre of Excellence for Gravitational Wave Discovery (OzGrav),
grant number CE170100004.
The authors thank Julian Carlin for enlightening conversations
about stochastic processes in general and the state-dependent
Poisson model of pulsar glitches in particular.
The data analysed in the paper are drawn from the Jodrell Bank
Observatory glitch catalog at 
{\tt http://www.jb.man.ac.uk/pulsar/glitches/gTable.html}.
Figures are prepared with the
``Cornerplot'' and ``Plot and compare histograms'' tools
in the MATLAB File Exchange,
authored by W. Adler and J. Lansey respectively.

\bibliographystyle{mn2e}
\bibliography{glitchstat-latest}

\appendix
\section{Supervised, iterative, numerical recipe for maximizing ${\cal L}$
 \label{sec:preappa}}
In this paper, we maximize the likelihood ${\cal L}$ defined by (\ref{eq:pre7})
with respect to the parameter set $\theta$ by executing the following,
iterative, step-by-step recipe,
which combines two species of automated numerical maximization algorithms
with human supervision based on screening the intermediate results.

(i) We set bounds on the parameters using a combination of the data,
order-of-magnitude estimates, and internal consistency constraints 
satisfied by (\ref{eq:pre1})--(\ref{eq:pre5}).
The bounds are specified in \S\ref{sec:pre3a}.

(ii) Working by hand, we run Monte Carlo simulations of
(\ref{eq:pre1})--(\ref{eq:pre5}) at a small number of random points in
the parameter space.
In each time series $X(t)$ generated thus,
we check whether or not certain key glitch features
(e.g.\ average size, maximum size, average waiting time,
latest epoch)
match the data approximately.
When one bound on a variable is finite and the other is infinite,
we start searching near the finite bound and gradually step further away. 
This part of the recipe is not systematic and relies on human discretion.

(iii) From the small set of {\em ad hoc} experiments in step (ii),
we record the parameter vector $\theta'$ with the highest ${\cal L}$.
This rudimentary estimate serves as a safety check on the
algorithmic searches to follow.
If the algorithms return a completely different parameter vector
with even higher ${\cal L}$, it counts as good news.
On the other hand, if the algorithms settle on a lower ${\cal L}$,
there are grounds for concern.
We also generate synthetic, Monte Carlo data for $\theta'$,
feed the data to the algorithms, and check if the algorithms return $\theta'$.

(iv) Starting and staying within the bounds established in step (i)
and explored in steps (ii) and (iii),
we run two automatic maximization algorithms,
particle swarm (PS) and Markov chain Monte Carlo (MCMC).
Details of the algorithms and their use are given in \S\ref{sec:pre3b}. 
As an example, for the three objects studied in this paper,
the supervised experiments in steps (ii) and (iii) yield the following bounds
for the automated maximisation algorithms: 
$\lambda_0 \in [0,1]$ (in units of ${\rm d}^{-1}$), 
$\tilde{t}_0 \in [\max_i(T_{i+1}-T_i),20\max_i(T_{i+1}-T_i)]$,
$\gamma^{-1} \tilde{X}_0 \in [\max_i(\Delta \nu_i),20\max_i(\Delta \nu_i)]$,
$\delta \in [0,2]$, 
$\beta \in [0,1]$,
and $\gamma \in [2\pi,30]$.
These bounds are used for all the PS searches. 
The more expensive MCMC searches are done over a smaller region, 
centred around the PS point estimates, 
in order to compute the marginalized likelihoods.

(v) We check the results from step (iv) against those from step (iii).
Optionally one can also run a brute force grid scan and a genetic algorithm
(for example) in the vicinity of the maxima returned in step (iv)
where necessary and appropriate. 
If any of these safety checks uncover a higher local maximum,
we return to step (iv) and run PS and MCMC again 
with a refined starting point.

\section{Epoch prediction with homogeneous and state-dependent Poisson processes
 \label{sec:preappb}}
In this appendix, we compare the predictive accuracy of 
(i) a homogeneous Poisson process with a constant rate equal to
the reciprocal of the arithmetic mean of the observed waiting times,
and (ii) a state-dependent Poisson process of the form specified
in \S\ref{sec:pre2},
whose variable rate is generated by the stress history $X(t)$
corresponding to the maximum likelihood estimates of the parameters
$\tilde{X}_0$, $\tilde{t}_0$, $\alpha$ (or equivalently $\lambda_0$),
$X(T_1^+)$, $\gamma$, $\beta$, and $\delta$.
The motivation for the comparative study is the results 
from \S\ref{sec:pre4} and \S\ref{sec:pre5},
which suggest that (ii) is a marginally better epoch predictor than (i).
The latter conclusion is tentative:
it is reached in only three pulsars and for
relatively small samples ($N\leq 35$)
and reverses if certain events are excluded,
e.g.\ the unusually large, penultimate glitch in PSR J0534$+$2200
(see Table \ref{tab:pre2}).
Below we employ Monte Carlo simulations
to study the matter further and obtain improved statistics.
Needless to say, the simulations cannot address the question
of whether (i) or (ii) is a better epoch predictor for a real pulsar;
this will be resolved in the future, as more astronomical data become available.
The simulations simplify verify the expected result that, 
if a state-dependent Poisson process is hypothetically at work in a pulsar,
then (ii) is a better epoch predictor than (i),
because (ii) is calibrated against the observed sizes and waiting times
and allows $\lambda[X(t)]$ to vary, 
while (i) is calibrated against the waiting times only
and holds $\lambda[X(t)]$ constant artificially.

\begin{figure}
\centering
\includegraphics[width=0.5\columnwidth]{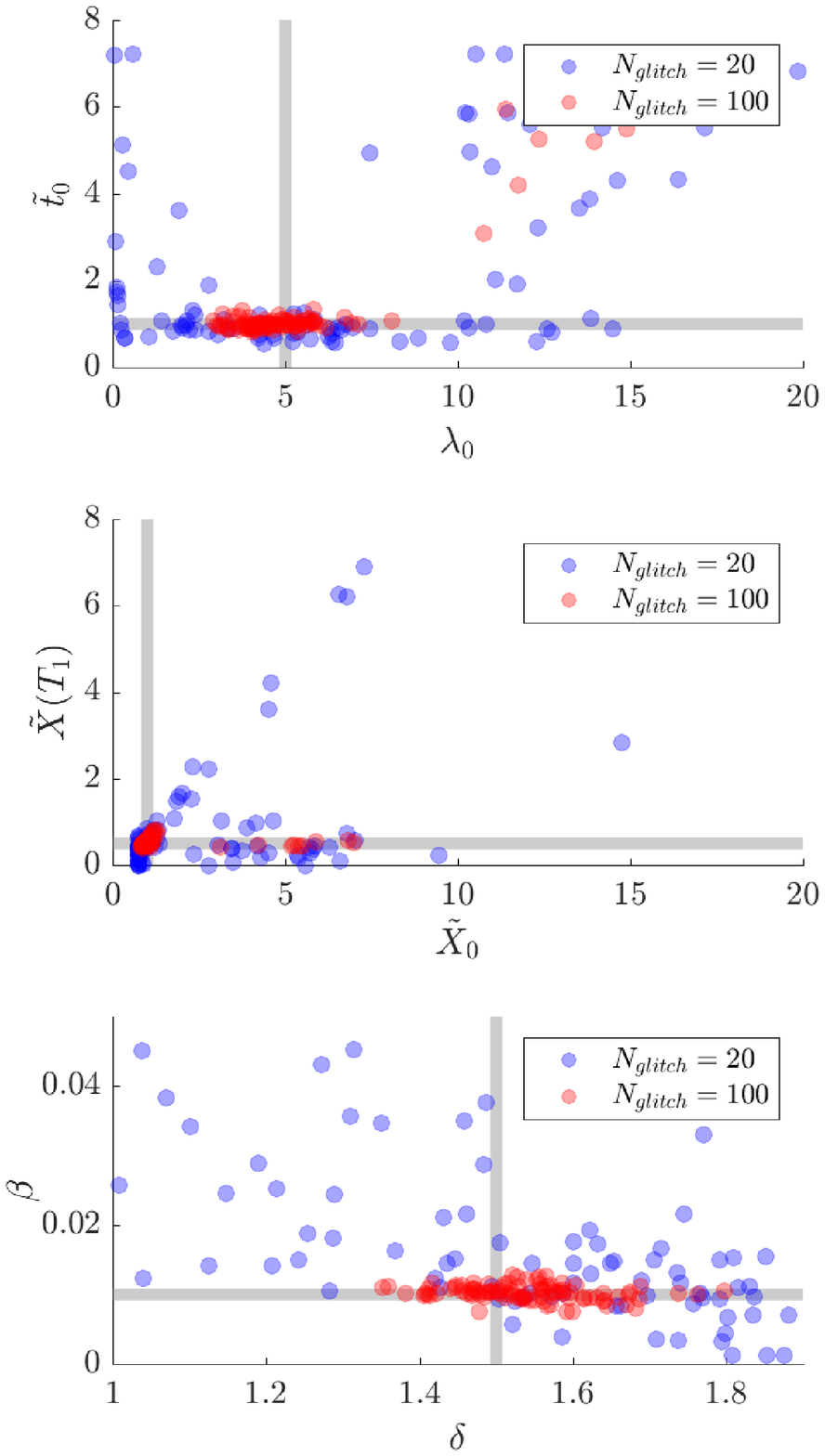}
\caption{
Dispersion of parameter estimates for PS searches of synthetic data 
generated by (\ref{eq:pre1})--(\ref{eq:pre5})
containing 20 glitches (shaded red circles; 96 realizations) 
and 100 glitches (shaded blue circles; 96 realizations),
visualized via three two-dimensional cross-sections of the parameter space:
$\tilde{t}_0$ versus $\lambda_0$ (top panel), 
$\tilde{X}(T_1)$ versus $\tilde{X}_0$ (middle panel),
and $\beta$ versus $\delta$ (bottom panel),
with $\gamma=2\pi$ fixed.
The grey shaded lines denote the true parameter values:
$\lambda_0=5$, $\tilde{t}_0=1$, 
$\tilde{X}_0=1$, $\tilde{X}(T_1^+)=0.5$, 
$\delta=1.5$, $\beta=0.01$.
}
\label{fig:pre10}
\end{figure}

\begin{figure}
\centering
\includegraphics[width=0.5\columnwidth]{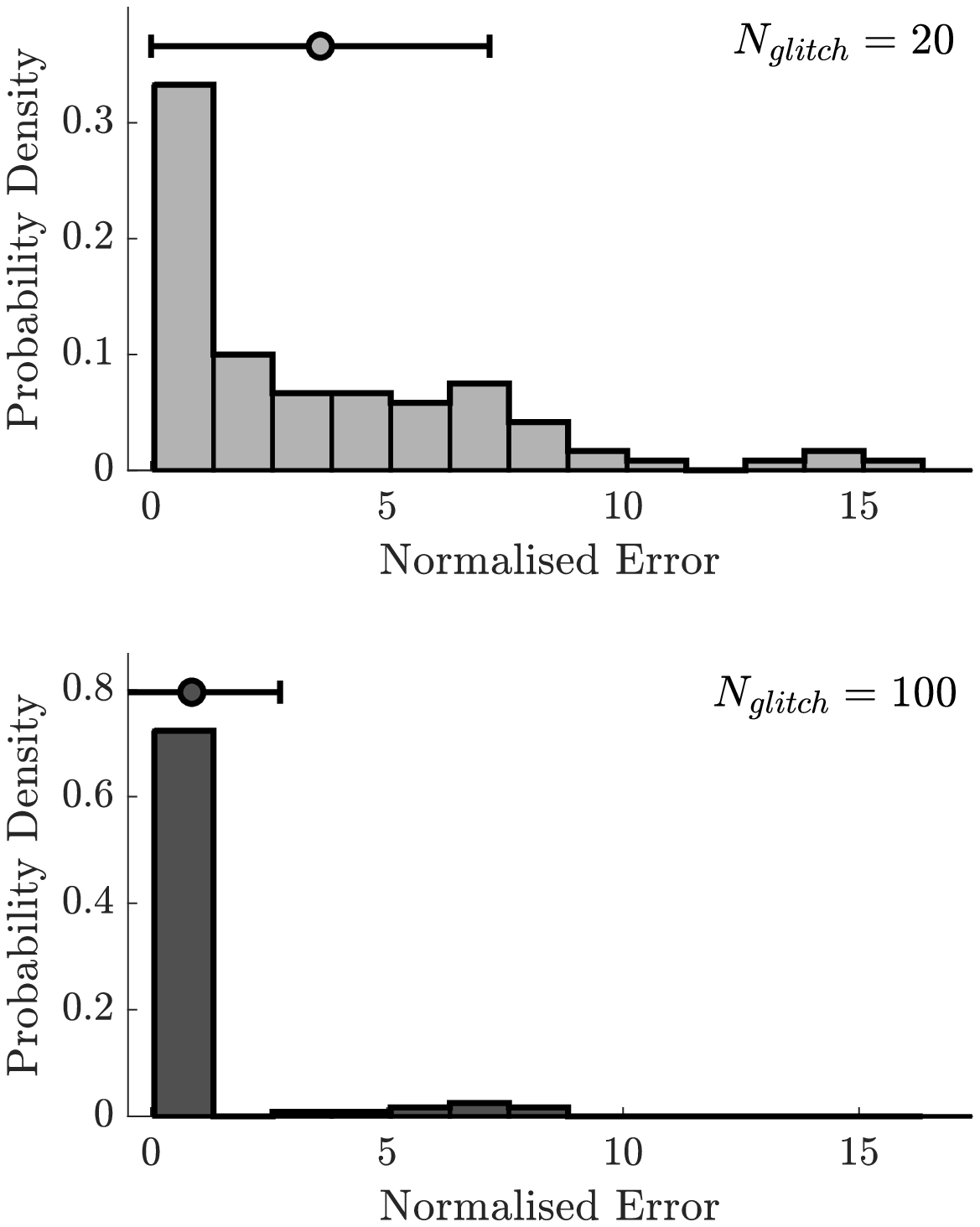}
\caption{
Normalized error between the true parameter values $\theta$ 
and PS estimates $\hat{\theta}$
for the Monte Carlo simulations in Figure \ref{fig:pre10}
for $N=20$ (top histogram; 96 realizations) 
and $N=100$ (bottom histogram; 96 realizations).
The normalized error is defined as
the six-dimensional Euclidean norm,
$[ \sum_{j=1}^6 (\theta_j - \hat{\theta}_j )^2 / \theta_j^2 ]^{1/2}$,
with $\gamma=2\pi$ fixed.
The mean and standard deviation are 3.59 and 3.59 for $N=20$
and $0.87$ and $1.87$ for $N=100$ respectively.
}
\label{fig:pre11}
\end{figure}

Figures \ref{fig:pre10} and \ref{fig:pre11} confirm that
the parameters of the state-dependent Poisson model are estimated with
increasing accuracy, as $N$ increases.
Equations (\ref{eq:pre1})--(\ref{eq:pre5}) are used to generate
$96\times 2$ realizations of $X(t)$ with $N=20$ and $N=100$
for $\theta$ as specified in the caption of Figure \ref{fig:pre10}.
The PS algorithm is applied to the synthetic data to produce 
maximum likelihood point estimates
of $\lambda_0$, $\tilde{t}_0$, $\tilde{X}_0$, $\tilde{X}(T_1^+)$,
$\beta$, and $\delta$.
(We fix $\gamma=2\pi$ as in \S\ref{sec:pre4a}--\S\ref{sec:pre4c}
to keep the computation tractable.)
The estimates are plotted pairwise to assist with visualization
in Figure \ref{fig:pre10}, 
with blue and red dots corresponding to $N=20$ and $N=100$ respectively.
One finds that the red dots are clustered more tightly around the true values
(grey horizontal and vertical lines) than the blue dots,
as expected.
Figure \ref{fig:pre11} makes the same point by presenting histograms
for the normalized distance between the true and estimated parameters,
as defined by the six-dimensional Euclidean norm.
One finds that
the $N=100$ histogram is narrower than the $N=20$ histogram;
its mean and standard deviation are $\lesssim 4$ times smaller.

\begin{figure}
\centering
\includegraphics[width=0.55\columnwidth]{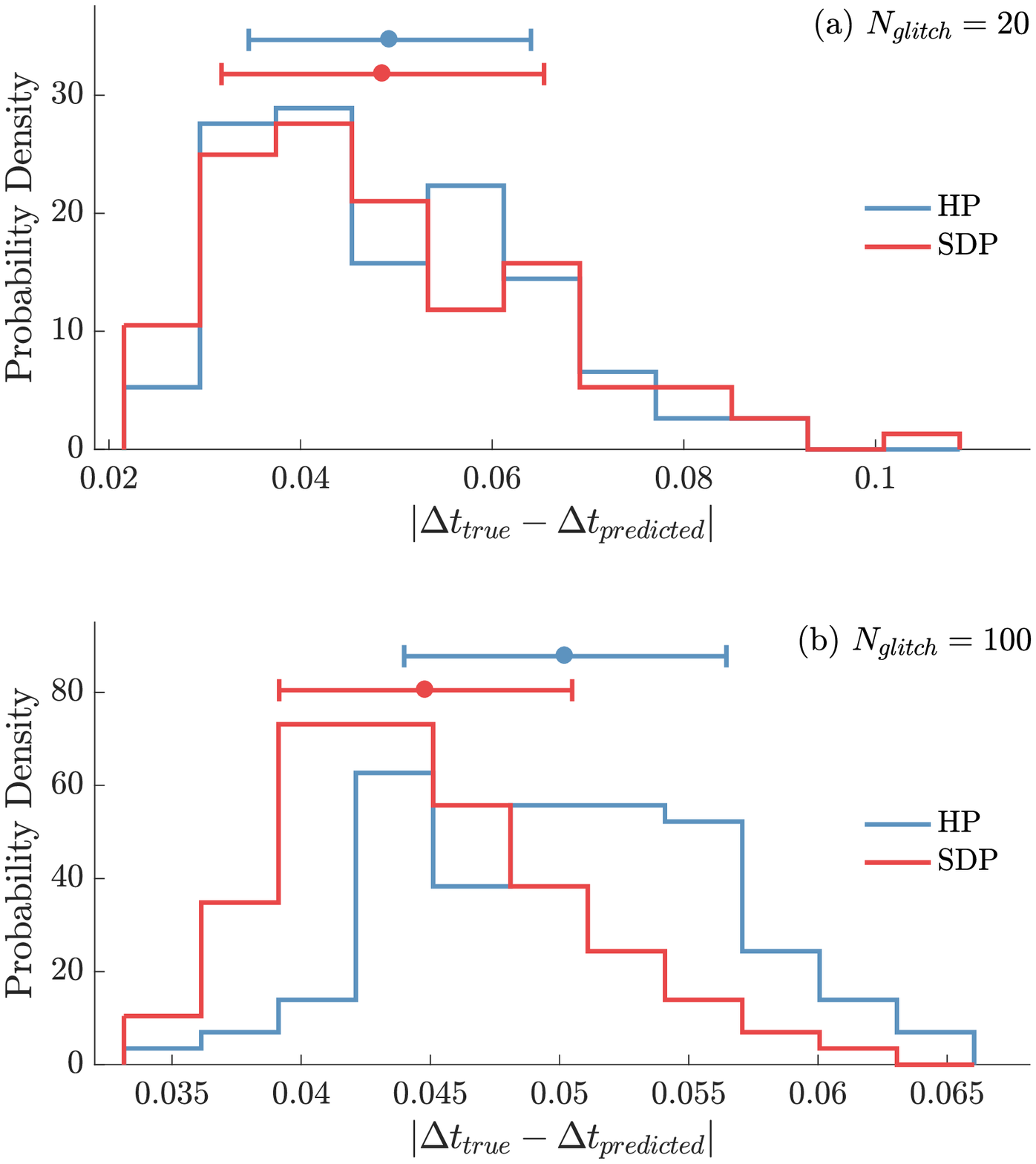}
\caption{
Histogram of the unsigned absolute error in the (dimensionless) predicted epoch, 
i.e.\ the predicted minus the true epoch
inferred from the state-dependent (red border)
and homogeneous (blue border) Poisson models,
for the Monte Carlo simulations in Figure \ref{fig:pre10}.
(a) $T_{2},\dots,T_{20}$.
(b) $T_2,\dots,T_{100}$.
The red and blue histograms overlap in some bins.
Color-coded dots and horizontal error bars indicate the mean
and standard deviation respectively for both models;
see also Table \ref{tab:pre4}.
}
\label{fig:pre12}
\end{figure}

Figure \ref{fig:pre12} confirms that epoch predictions 
based on a state-dependent Poisson model
are more accurate than epoch predictions
based on a homogeneous Poisson model,
if a state-dependent Poisson process governs the underlying dynamics.
Although by itself this is not surprising,
the purpose of Figure \ref{fig:pre12} is to quantify roughly how many events
are needed, before the advantage asserts itself clearly.
Histograms are presented of the unsigned absolute error 
in the predicted epochs $T_2,\dots,T_N$ for $N=20$ and $N=100$,
computed for $96\times 2$ realizations of $X(t)$ for a state-dependent 
Poisson process with the same parameters as in Figure \ref{fig:pre10}.
The state-dependent and homogeneous Poisson models are equally accurate
for $N=20$, but the former outperforms the latter for $N=100$,
where its mean error is 11 per cent smaller;
see Table \ref{tab:pre4}.
The advantage is modest, for the reasons expressed in \S\ref{sec:pre5a}.
The Monte Carlo simulations can be extended to study how the error
in predicting $T_{N+1}$ scales with $N$,
when warranted by additional data.
Again we emphasize that the Monte Carlo results say nothing about whether
glitch activity in real pulsars obeys a state-dependent Poisson process. 

\begin{table}
\begin{center}
\begin{tabular}{rcccc}\hline
 $N$ & Mean (SDP) & Std dev (SDP) & Mean (HP) & Std dev (HP) \\ \hline
 20 & 0.0486 & 0.0422 & 0.0493 & 0.0414 \\
 100 & 0.0448 &  0.0420 & 0.0502 & 0.0461 \\
 \hline
\end{tabular}
\end{center}
\caption{
Means and standard deviations of the (dimensionless)
epoch prediction error histograms 
in Figure \ref{fig:pre12}
versus the number of glitches $N$
for the state-dependent (SDP) and homogeneous (HP) Poisson models.
}
\label{tab:pre4}
\end{table}

\end{document}